\documentclass[prl,twocolumn,superscriptaddress]{revtex4-2}

\usepackage{graphicx,color}
\usepackage{amsmath}
\usepackage{amssymb}
\usepackage{natbib}
\usepackage{tabularx}
\usepackage{multirow}
\usepackage[colorlinks=true,linkcolor=blue,citecolor=blue,urlcolor=blue]{hyperref}
\usepackage{ulem}
\usepackage{placeins}

\newcommand{\dto}{Dy$_2$Ti$_2$O$_7$}
\newcommand{\hto}{Ho$_2$Ti$_2$O$_7$}

\begin{document}

\title{Satisfaction and Violation of the Fluctuation-Dissipation Relation in spin ice materials}

\author{F. Morineau} 
\email[]{felix.morineau@neel.cnrs.fr}
\affiliation{Institut N\'eel, CNRS, Universit\'e Grenoble Alpes, 38042 Grenoble, France}
\author{V. Cathelin} 
\affiliation{Institut N\'eel, CNRS, Universit\'e Grenoble Alpes, 38042 Grenoble, France}
\author{P. C. W.  Holdsworth} 
\affiliation{ENS de Lyon, CNRS, Laboratoire de Physique, F-69342 Lyon, France}
\author{S. R. Giblin} \affiliation{School of Physics and Astronomy, Cardiff University, Cardiff, CF24 3AA, United Kingdom}
\author{G. Balakhrishnan} \affiliation{Department of Physics, University of Warwick, Coventry, CV4 7AL, United Kingdom}
\author{K. Matsuhira}\affiliation{Kyushu Institute of Technology, Kitakyushu 804-8550, Japan}
\author{C. Paulsen} 
\affiliation{Institut N\'eel, CNRS, Universit\'e Grenoble Alpes, 38042 Grenoble, France}
\author{E. Lhotel}
\email[]{elsa.lhotel@neel.cnrs.fr}
\affiliation{Institut N\'eel, CNRS, Universit\'e Grenoble Alpes, 38042 Grenoble, France}

\begin{abstract}
We test the fluctuation-dissipation relation (FDR) in spin ice materials \dto\ and \hto\ by measuring both the magnetic noise and the out-of-phase part of the susceptibility and comparing their ratio. We show that it is satisfied at temperatures well into the nonergodic region below 600 mK, indicating local equilibrium. In both materials, below 400 mK, low frequency violations develop, showing an excess of noise as in spin glasses, with a frequency threshold of 0.1 Hz. New relaxation pathways and aging properties are unveiled in this frequency range in the ac susceptibility. The FDR remains valid at higher frequencies down to 150 mK.

\end{abstract}

\maketitle
At thermodynamic equilibrium, the fluctuation-dissipation theorem relates the spontaneous fluctuations in a system and the linear response function through the fluctuation-dissipation relation (FDR) \cite{Kubo66}.  When the system is not at equilibrium, deviations from the fluctuation-dissipation theorem are expected and were indeed reported in the context of  glasses and liquid crystals \cite{Grigera99, Bellon01, Joubaud09, Oukris10}.  A comprehensive theoretical description of the relation between fluctuations and dissipation in out-of-equilibrium systems nevertheless remains challenging~\cite{Sarracino19}. 

In the specific context of magnetism, out-of-equilibrium properties have been mainly studied in spin-glasses \cite{Herisson02} and more recently in superspin-glasses (assemblies of interacting nanoparticles) \cite{Komatsu11}. In these systems, simultaneous measurements of magnetic fluctuations and the response function have been performed to directly probe the FDR. They show that these systems can be described by an effective temperature different from the thermal bath which quantifies the deviations from the equilibrium case. In this description, both the fluctuations and response function are made of two contributions: quasiequilibrium or staggered, which respects time translation invariance, and aging, which represents the fact that the response and correlation functions decay slowly with time. This hypothesis, corresponding to a weak ergodicity breakdown, can be understood by considering that the system needs an infinite time to reach its equilibrium state, but that locally its behavior is governed by equilibrium dynamics~\cite{Cugliandolo11}. 
An interesting system where out-of-equilibrium properties are reported and which may break the fluctuation-dissipation relation is spin ice \cite{Jaubert_book,Raban22}. The spin ice state was originally observed in pyrochlore oxide compounds $R_2M_2$O$_7$,  where the $R$ and $M$ ions lie on two interpenetrated corner sharing tetrahedron lattices, when $R=$Ho, Dy and $M$ is a non-magnetic element. It is characterized by the ``Pauling states,'' satisfying the ice rule with two spins pointing inward and two outward on each tetrahedron, stabilized experimentally below about 1 K. 
The Pauling states form a narrow quasidegenerate band, stabilized by geometrical frustration in the absence of structural disorder, retaining extensive entropy down to the lowest temperatures. Below 650 mK magnetization measurements show a clear bifurcation between the field-cooled and zero field-cooled curves \cite{Snyder04, Petrenko11}, which indicates an ergodicity breaking and suggests that the system enters into an out-of-equilibrium state. This freezing results from the strong slowing down of the dynamics at these temperatures, due to the rarefaction of the excitations in the system, the so-called magnetic monopoles \cite{Castelnovo08} (which correspond to a violation of the local ice rule, generating a ``3-in, 1-out" or ``3-out, 1-in" configuration in a tetrahedron). In addition, the system never reaches the predicted ordered ground state at very low temperature \cite{Melko04}, even after months of waiting \cite{Giblin18, Samarakoon22b}. 

In this Letter, we address the nature of this low temperature state by probing the FDR in two spin ice systems \dto\ and \hto. In both compounds, we identify three characteristic regions: (i) the high temperature region, where the system is trivially in equilibrium, and the FDR is obeyed, as expected; (ii) the intermediate region, where ergodicity breaking is observed in the magnetization, but where we find that the FDR remains obeyed indicating that local equilibrium is preserved; (iii) the low temperature region, where the FDR is violated at low frequency, typically below 0.1 Hz, indicating that the system enters an out-of-equilibrium state. Furthermore, we observe a dissipation process that was not reported to date, as well as aging effects.

The FDR in a magnetic system relates the spectral noise density $S(f)$ to the dissipative part of the ac magnetic susceptibility $\chi''(f)$. In the case of a stationary system $S(f)$ corresponds to the Fourier transformed autocorrelation function of the magnetization  [$\overline{M^2(f)}$] and is characteristic of the spontaneous fluctuations. The relation is written as \cite{Refregier87}
\begin{gather}
S(f)=D(f),  \label{eq_FDT} \\
 \textrm{\quad where \ } S(f)=\overline{M^2(f)} \textrm{\quad and \ } D(f)=\frac{2k_{\rm B}T}{\pi V} \frac{\chi''(f)}{f}, \nonumber
\end{gather}
where $T$ is the temperature, $f$ is the frequency, $k_{\rm B}$ is the Boltzmann constant, and $V$ is the sample volume. Satisfaction of the FDR indicates that the system is at least in local equilibrium, such that the exchange of energy to and from the heat bath is balanced. Its violation ensures an imbalance in this exchange which generates thermodynamic forces driving temporal evolution through configuration space and aging.
Such violations are parameterised by the fluctuation dissipation ratio $S(f)/D(f)$, characteristic of an effective temperature, equal to the ratio times the temperature of the heat bath \cite{Cugliandolo11}. In the aging regime of spin glasses, the effective temperature is observed to be greater than that of the heat bath \cite{Herisson02}.

To probe this relation, we have developed a superconducting quantum interference device magnetometer equipped with a dilution refrigerator, which allows us to measure the spectral noise density and the ac susceptibility of a zero field cooled sample in the same series of measurements in order to compare both unambiguously (See Supplementary material \cite{supmat}).
The experiments were carried out down to 150 mK and 163~mK, respectively on two single crystals: a parallelepiped sample of $^{162}$Dy$_2$Ti$_2$O$_7$ of $3.7 \times 5.1 \times 14$ mm$^3$ (with zero Dy nuclear moment) \cite{Fennell04,Giblin18} and a nearly cylindrical \hto\ sample of 4.8~mm diameter and 8.9~mm length \cite{supmat}.

\begin{figure}
\includegraphics[width=7.7cm]{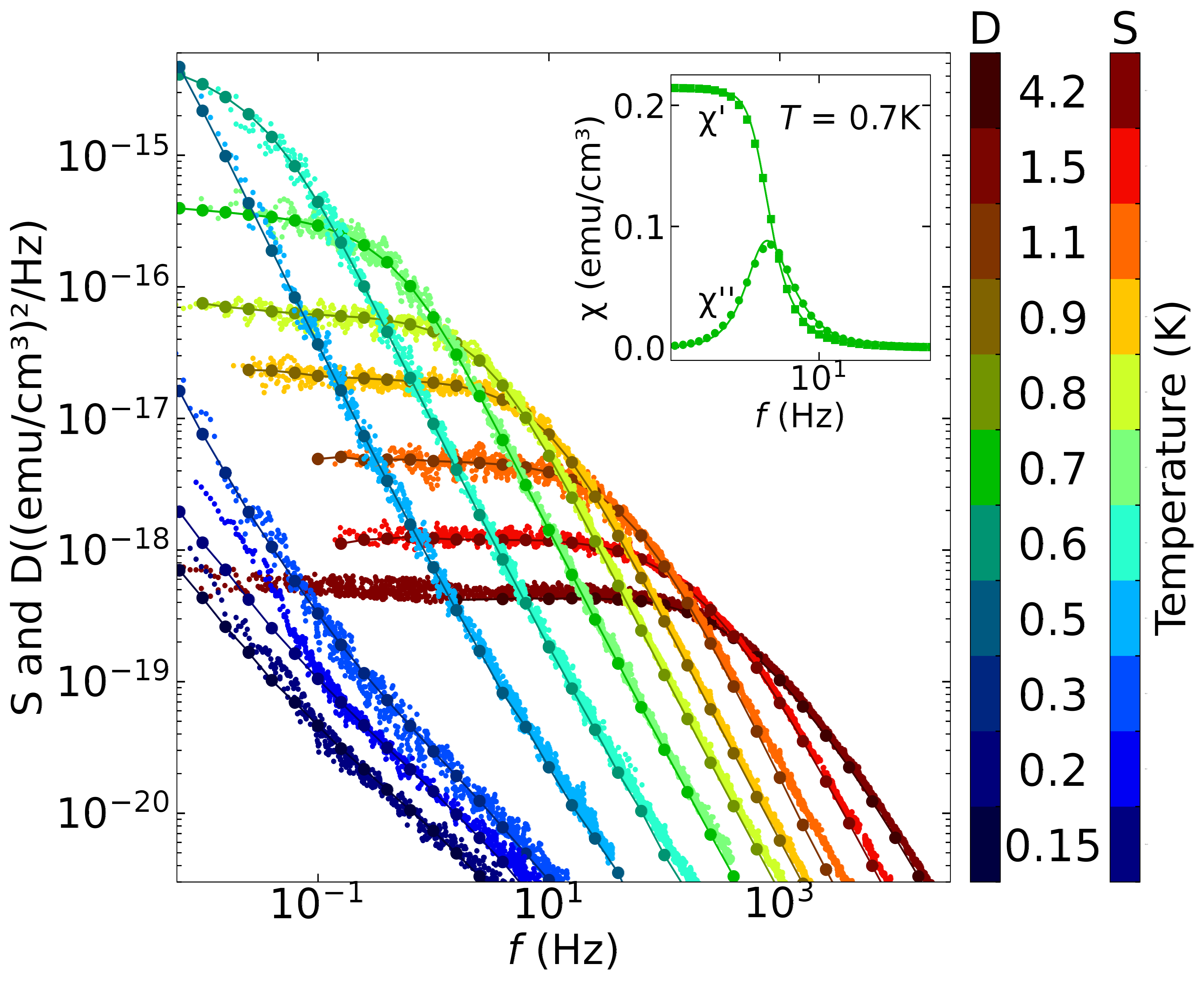}
\caption{\label{fig_FDT_DTO} FDR plot for \dto\ on a logarithmic scale: $S(f)$ (small dots) and $D(f)$ (big dots) measured between 4.2 K and 150 mK. Lines are guides to the eye. Inset: ac susceptibility $\chi'$ and $\chi''$ vs $f$ measured at 700 mK. The solid lines show the fit to the Cole-Davidson equation (\ref{Davidson_Cole}) with $\tau=0.429$ s, $\beta=0.67$, $\chi_S=0$ and $\chi_T=0.213$ emu.cm$^{-3}$. }
\end{figure}

\begin{figure}[h!]
\includegraphics[width=7.7cm]{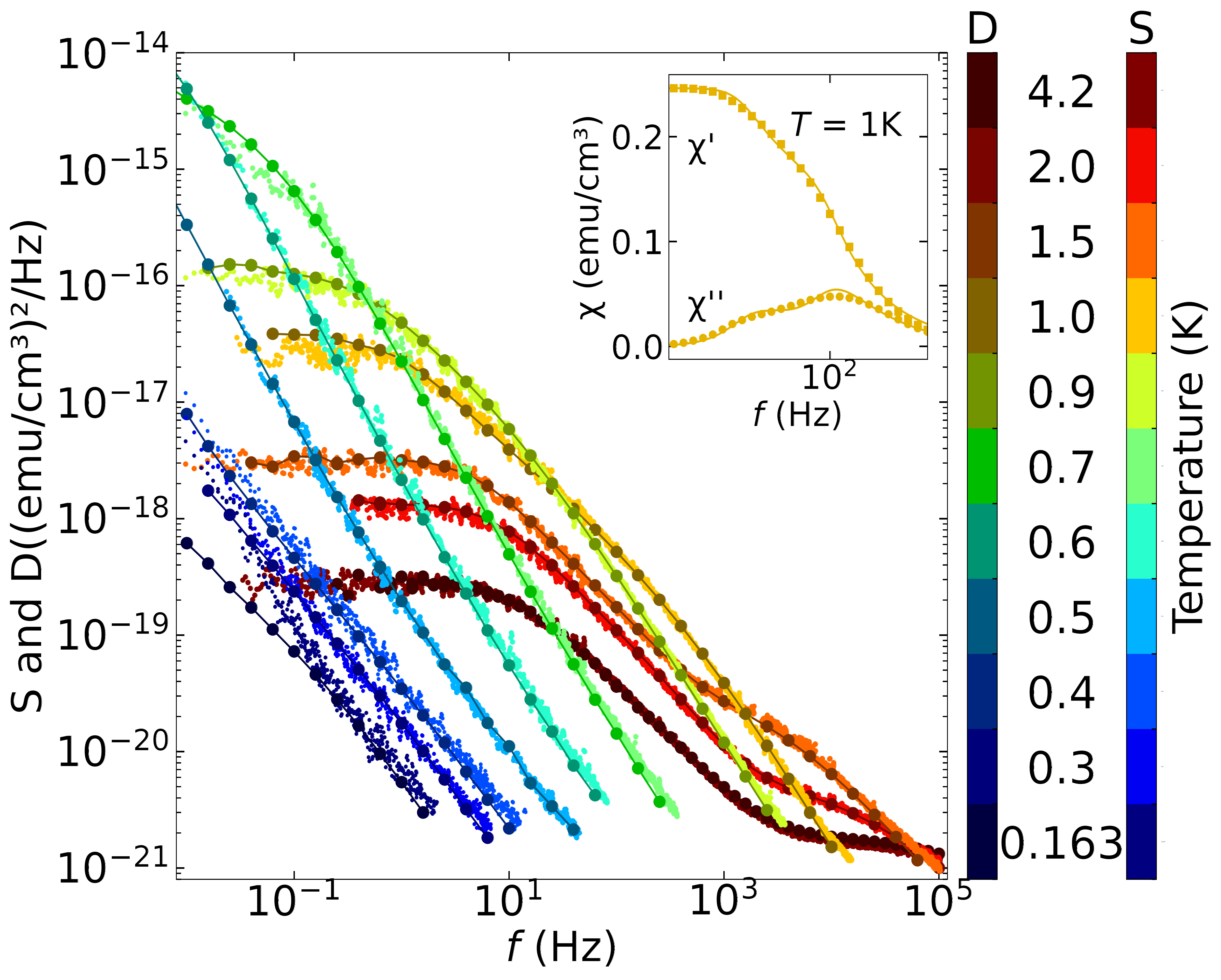}
\caption{\label{fig_FDT_HTO} FDR plot for \hto\ on a logarithmic scale: $S(f)$ (small dots) and $D(f)$ (big dots) measured between 4.2~K and 163 mK. Lines are guides to the eye. Inset: ac susceptibility $\chi'$ and $\chi''$ vs $f$ measured at 1 K.  The solid lines show the fit to a sum of two Cole-Davidson equations (\ref{Davidson_Cole}) with $\tau_1=0.098$~s, $\beta_1=0.31$, $\chi_{T1}=0.116$ emu.cm$^{-3}$, $\tau_2=0.002$ s, $\beta_2=0.41$, $\chi_{T2}=0.13$ emu.cm$^{-3}$ and $\chi_{S1}=\chi_{S2}=0$. }
\end{figure}

In Figs. \ref{fig_FDT_DTO} and \ref{fig_FDT_HTO}, the spectral noise density $S(f)$ is plotted together with $D(f)$ for \dto\ and for \hto, respectively. In \dto, the measured noise is consistent with previous measurements \cite{Dusad19, Samarakoon22, footnote}: above 0.6 K, $S(f)$ follows a power law dependence at high frequency and exhibits a plateau at low frequency; the separation between the two regimes defining the characteristic relaxation time. At lower temperature, the plateau regime is not reached in the measurements. In \hto, the shape of the noise is slightly different with the presence of two shoulders down to 1 K, due to the existence of two relaxation times in the system \cite{Wang21}, which can be clearly seen in the ac susceptibility (see inset of Fig.~\ref{fig_FDT_HTO}). Despite these differences, in both systems, $S(f)$ and $D(f)$ overlap down to 300 and 400~mK, in \dto\ and \hto\ respectively, thus well below the freezing temperature (650~mK) measured in magnetization measurements. 
This means that the FDR [Eq.~(\ref{eq_FDT})] is obeyed even in the non-ergodic regime and suggests that, while the system fails to evolve towards its equilibrium state when lowering the temperature or applying a magnetic field, the local dynamics of the spin ice state behave as if the system was at equilibrium. 
This is in agreement with simulations  on the dumbbell model of spin ice parameterized for \dto. Using stochastic dynamics, it was found that the FDR is obeyed down to $400$ mK, when the system is allowed to reach local equilibrium \cite{Raban22}. 
This contrasts, however, with the picture of spin ice behaving as a glass at low temperature \cite{Eyvazov18, Samarakoon22b}. 
This implies that for all intents and purposes, down to these temperatures, the ac susceptibility and noise measurements are equivalent methods to probe the dynamics and correlation functions in the system. Nevertheless, the ease and precision of ac susceptibility measurements make it the better tool. 

\FloatBarrier

\begin{figure}[h!]
\includegraphics[width=8.5cm]{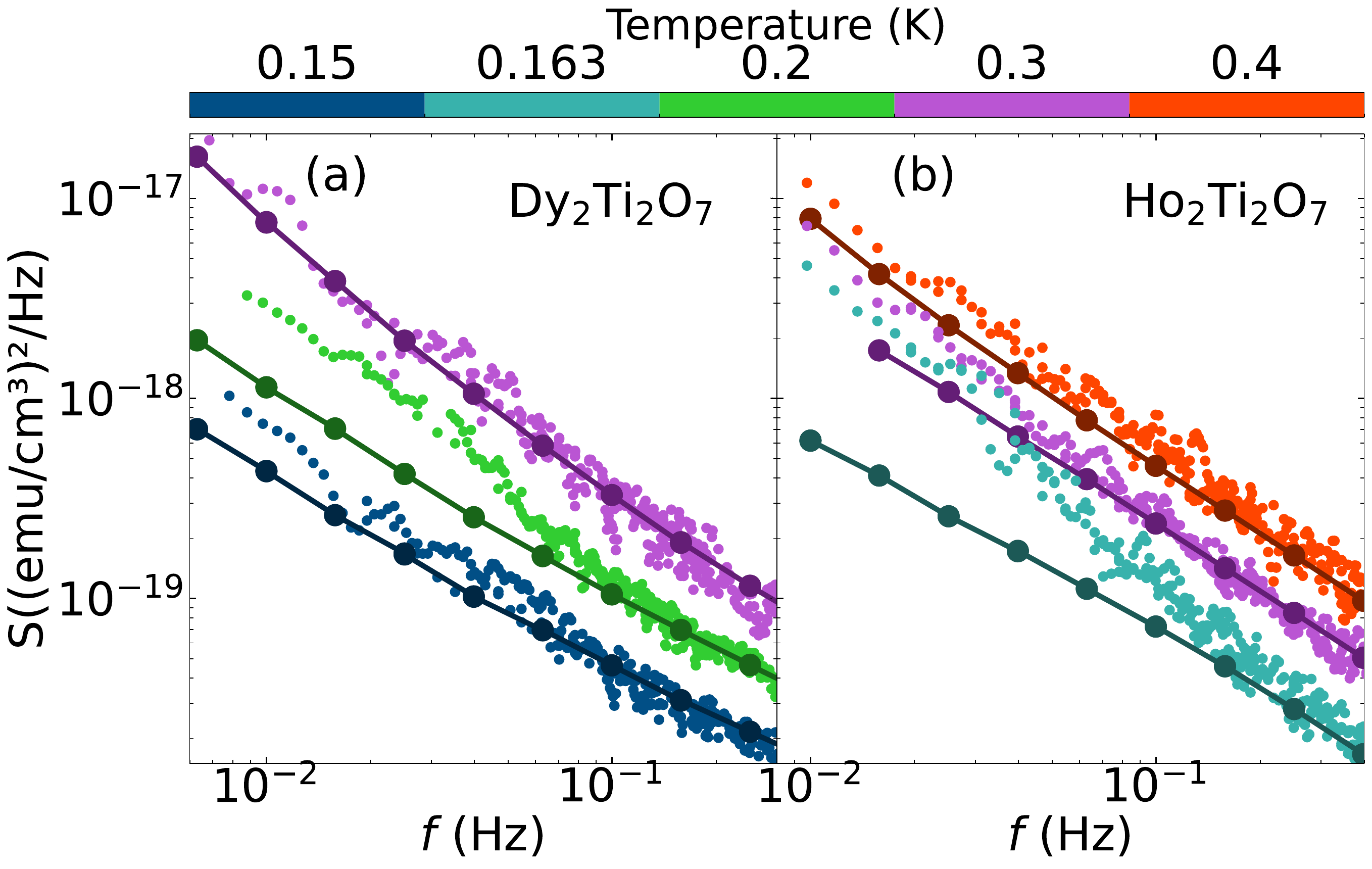}
\caption{\label{fig_LT-LF} FDR plot for (a) \dto\ and (b) \hto\ plotted on a logarithmic scale at low temperature and low frequency: $S(f)$ (small dots) and $D(f)$ (big dots). (c) Fluctuation dissipation ratio $S(f)/D(f)$ calculated below 0.016~Hz for \dto\ (in blue) and \hto\ (in gold) from base temperature to 800~mK. The solid black line corresponds to a ratio equal to 1 which is expected when there is no violation of the FDR.}
\end{figure}

Below 300 mK for \dto\ and 400 mK for \hto\ deep in the nonergodic regime, an excess noise is observed in $S(f)$ compared to $D(f)$ at the lowest measured frequencies, typically below 0.1 Hz [see Figs.~\ref{fig_LT-LF}(a) and \ref{fig_LT-LF}(b)]. The value of this excess noise is much larger than the noise measured at higher temperatures and in the empty  sample holder experiment, which makes us confident that it is intrinsic to the physics of spin ice. The excess noise is associated with  an abrupt increase of the $S(f)$ slope in the low frequency regime, while the dissipation part $D(f)$ obtained from the susceptibility increases smoothly.
The difference between the fluctuation and the dissipation contributions  indicate that the FDR is violated in this low temperature and low frequency regime for both  \dto\ and \hto.  In \dto, the fluctuation dissipation ratio reaches about 2 below 0.016 Hz and 200 mK. The amplitude of the violation is stronger in \hto\ where the fluctuation dissipation ratio reaches about 6 at 163~mK [see Fig.~\ref{fig_LT-LF}(c)]. This result means that, although the absolute scale is different, an equilibrium system would require a temperature of approximately one kelvin to achieve this ratio \cite{Cugliandolo11}.

This scenario is characteristic of out-of-equilibrium systems where multiple time scales are present \cite{Cugliandolo97} and has been observed in spin-glasses \cite{Herisson02}. In the high frequency regime, above the threshold of approximately 0.1~Hz, the FDR is satisfied, indicating local equilibrium over short time and length scales. The breakdown of the FDR with high effective temperature indicates the onset of non-ergodicity over mesoscopic time and length scales with consequent temporal evolution and aging. In \hto, where the effect is stronger, a small temperature dependence is observed in the crossover frequency which moves from 0.05 to 0.16 Hz between 400 and 163~mK. 

\begin{figure}
\includegraphics[width=8.5cm]{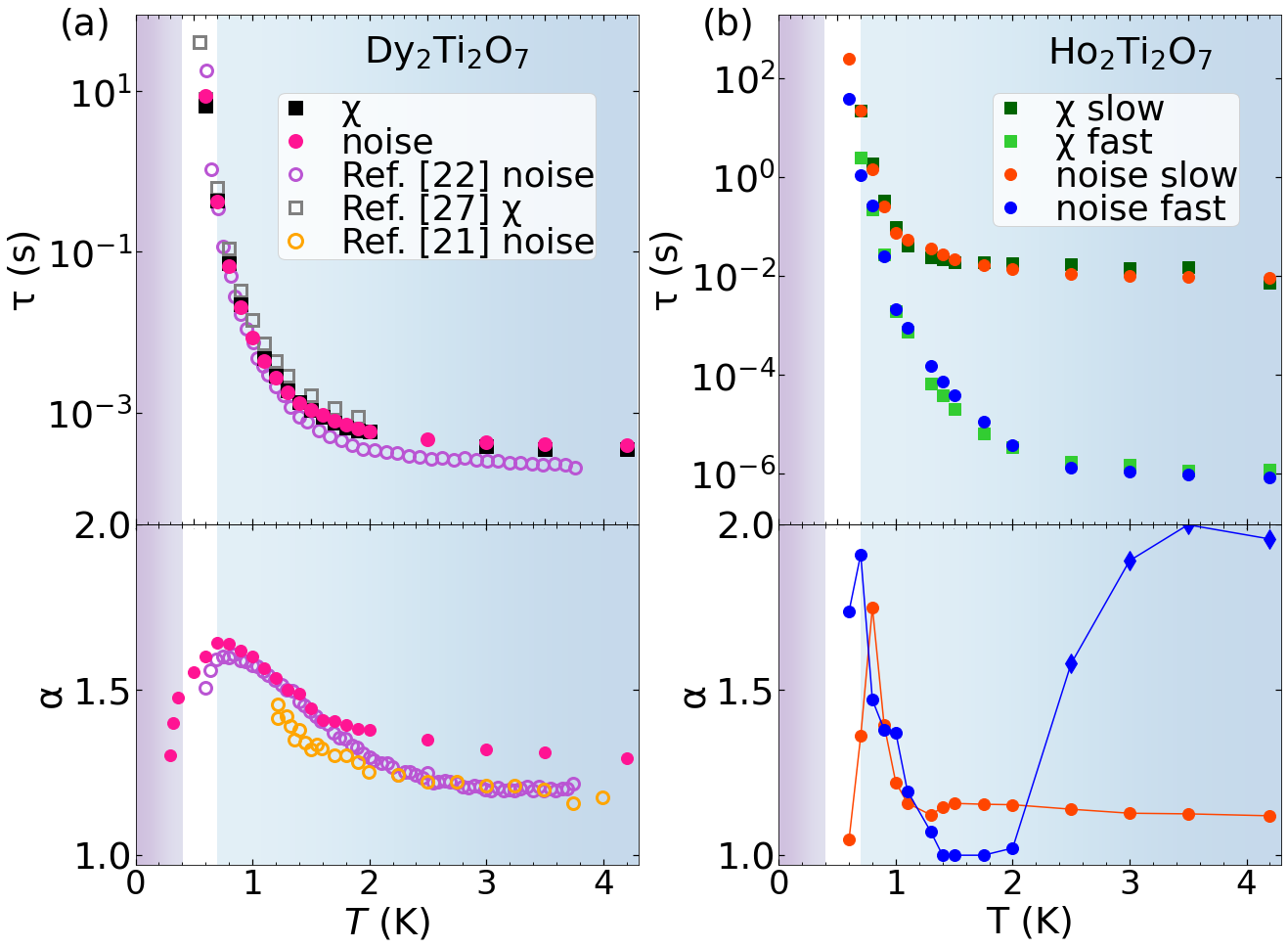}
\caption{\label{tau_alpha} (a) Relaxation time $\tau$ (top panel) and exponent $\alpha$ (bottom panel) in \dto\ obtained from the fits of the noise data using Equation (\ref{fit_S}), together with the $\tau$ obtained from the fits of $\chi(f)$ with Equation (\ref{Davidson_Cole}) and the results from Refs. \onlinecite{Samarakoon22, Matsuhira11, Dusad19}. Below 600 mK, it is not possible to fit the noise and ac susceptibility due to the absence of shoulder in the noise and of maxima in $\chi''$. The $\alpha$ parameter can nevertheless be obtained from the slope of the noise. (b) Relaxation times $\tau_{\rm slow}$ and $\tau_{\rm fast}$ (top panel) and exponents $\alpha_{\rm fast}$ and $\alpha_{\rm slow}$ (bottom panel) in \hto\ obtained from noise measurements. Above 2 K, $\alpha_{\rm fast}$ is obtained with a high uncertainty because the power law slope of the mode is at the limit of the frequency window (diamond symbols). The relaxation times obtained from ac susceptibility are also shown. The shaded backgrounds indicate the three temperature regimes: (i) equilibrium (blue), (ii) local equilibrium (white) and (iii) out of equilibrium (purple).}
\end{figure}

We have analyzed the detailed functional form of $S(f)$ and the ac susceptibility $\chi(f)$, extracting the characteristic relaxation time $\tau$ and exponents $\alpha$ and $\beta$ from expressions
\begin{align}
S(f)&=\frac{S_0}{1+(2\pi f \tau)^{\alpha}}, \label{fit_S} \\
\mathrm{and} \quad \chi(f)&=\chi_{\rm S}+\dfrac{\chi_{\rm T}-\chi_{\rm S}}{(1+i2\pi f \tau)^{\beta}} \label{Davidson_Cole}
\end{align}
\cite{Dusad19, Wang21}. $S_0$ is the noise at zero frequency, $\chi_S$ and $\chi_T$ are the adiabatic and isothermal susceptibilities respectively. As in previous experiments \cite{Matsuhira11, Quilliam11, Yaraskavitch12, Wang21, Samarakoon22}, we find (see Fig.~\ref{tau_alpha}) that the time scale $\tau$ diverges below 1 K but clearly above the temperature scale on which FDR violations develop. Hence, while the divergence of $\tau$ is compatible with the loss of global ergodicity and the zero field-cooled—field-cooled splitting for the static susceptibility it does not appear to be relevant for the onset of FDR violations. 
 
For \dto, three dynamical regimes can clearly be observed from the temperature dependence of the $\alpha$ parameter (or $\beta$ for the susceptibility) \cite{Samarakoon22}. Above 2~K and at least up to 4.2 K, when the system is not yet in the spin ice state, $\alpha$ is almost constant. It nevertheless remains below 2, the value for a random walk process, which would be expected for paramagnetic magnetic moments with a single relaxation time. When entering the spin ice regime, $\alpha$ increases, reaching about 1.6 and revealing correlated, constrained dynamics \cite{Hallen22}. When entering the freezing regime below 700 mK, $\alpha$ decreases once again reaching 1.3 at 300 mK [see Fig.~\ref{tau_alpha}(a)]. These results further illustrate that, while FDR violations surely require complex correlated dynamics, the reverse is not true: $\alpha$ differs from two over a large range of temperatures where the  FDR is satisfied indicating global or local equilibrium.

The analysis for \hto\ turns out to be less straightfoward due to the presence of two relaxation times \cite{Wang21}. Using the noise, as well as the in-phase and out-of-phase susceptibility we could determine the relaxation times and $\alpha$ parameters [see Fig.~\ref{tau_alpha}(b)] by assuming a sum of two independent contributions following Eqs.~(\ref{fit_S}) for $S(f)$ and (\ref{Davidson_Cole}) for $\chi(f)$. Compared to previous analysis with ac susceptibility only \cite{Wang21}, the combination of the two sets of data allows us to deconvolute the contributions of the two relaxation times, especially at low frequency. We find that they have a similar temperature dependence, their relative intensity being almost constant as a function of temperature (see Supplemental Material~\cite{supmat}). These two time scales are thus probably related to the existence of two single spin flip tunneling times resulting from the local spin configurations as proposed theoretically \cite{Tomasello15,Tomasello19}.
Qualitatively, the relaxation times have a similar temperature dependence to those in \dto: they have a nearly constant value above 2.5 K, which corresponds to the intrinsic tunnelling time and strongly increase at low temperature when entering the spin ice regime where slow monopole dynamics are present. The three regimes are also observed in the $\alpha$ parameters, especially the drop at low temperature below 700 mK. 

\begin{figure}
\includegraphics[width=8cm]{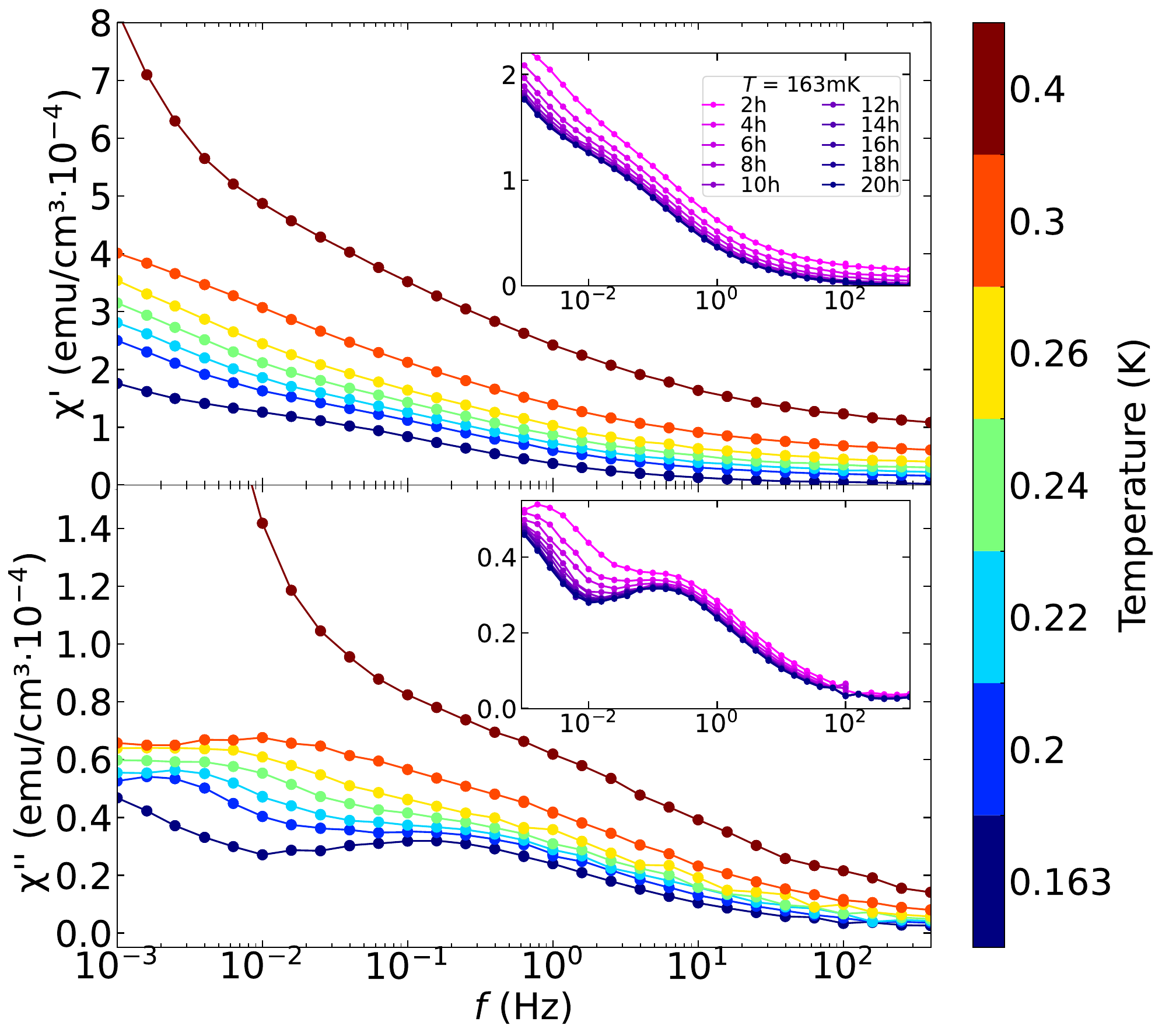}
\caption{\label{fig_Xac_LT} ac susceptibility $\chi'$ (top panel) and $\chi''$ (bottom panel) vs $f$ for different temperatures below 300 mK in \hto. Inserts: Susceptibility measured at 163 mK after different waiting times.}
\end{figure}

In order to better understand the FDR violation we have studied the ac susceptibility below 400~mK, where the signal is very small.
Interestingly, our ac susceptibility data reveal a relaxation phenomenon that appears at very low temperature and is different from previously identified processes \cite{Matsuhira11, Quilliam11, Yaraskavitch12, Wang21}. It manifests itself as two peaks in $\chi"$ of very small amplitude, which we could observe thanks to a greatly improved sensitivity (see Fig.~\ref{fig_Xac_LT}).
The peaks appear in both \dto\ and \hto\ samples, at slightly different frequencies~\cite{supmat}. They are not directly visible in the noise, but when converting $S(f)$ into $\chi"(f)$ using the FDR relation, the highest frequency peak, which sits in the frequency range where the relation is satisfied can just be resolved. However, the peak at the lowest frequency (below 0.1 Hz) is not visible, due to the presence of excess noise \cite{supmat}.
This low intensity signal shown in Fig.~\ref{fig_Xac_LT} has faster characteristic times (typically a few seconds compared to a few hundreds of seconds for the processes described above) with little temperature dependence compared to the majority relaxation processes. It is thus clearly associated with a new relaxation mechanism, which appears masked at higher temperature, becoming visible only when the other relaxation channels are too slow.  
In addition to these relaxation processes, ac susceptibility exhibits aging precisely in the regime where the FDR breakdown occurs (see inset of Fig.~\ref{fig_Xac_LT}) in close analogy with spin glasses in the same regime \cite{Herisson02}.

The FDR violations and aging could be due to the presence of trapped noncontractible monopole pairs \cite{Raban22} which can drive the system out of equilibrium \cite{Castelnovo10,Paulsen14, Paulsen16, Paulsen19}, although this seems unlikely as such pairs are only short-lived at 300 mK \cite{Castelnovo10,Raban22}  and require rapid thermal quenches to be stabilized at lower temperatures. However, it is possible that the spin ice materials begin to be sensitive to the degeneracy lifting of the Pauling states induced by the dipole interactions \cite{Melko04,Isakov05} in this temperature range. This finite energy bandwidth, which is a correction to the monopole picture of spin ice could lead to a rough free energy surface that drives FDR violations, as in spin-glasses \cite{Bouchaud98}. The emerging short timescale could then correspond to monopoles being confined in short loops, with the long timescale a deconfinement time. This separation of scales would be explicitly due to finite  bandwith of energy barriers between Pauling states \cite{Samarakoon22} which is absent in the idealised monopole picture~\cite{Raban22}. These ideas could be tested but the investigation of aging effects in the experimentally observed temperature range would require challenging simulations using the full dipolar spin-ice Hamiltonian \cite{Samarakoon22b,Samarakoon22}.


An important remark concerns the role of demagnetizing effects. Because of the ferromagnetic correlations in spin ice, corrections to the susceptibility due to demagnetizing effects can be very important \cite{Bovo13, Quilliam11}. When the real part of the ac susceptibility is sizable, such corrections mix the in-phase and out-of-phase parts of the susceptibility, resulting in a larger intrinsic $\chi_i''$ that is shifted to higher frequencies compared to measured data. However, in our experimental data of Figs.~\ref{fig_FDT_DTO} and \ref{fig_FDT_HTO}, the FDR is obeyed when comparing $S(f)$ with $D(f)$ which has not been corrected for demagnetizing effects. This seems counterintuitive as noise fluctuations are measured in a zero field environment \cite{supmat}, but presumably fluctuations do create their own internal fields which must also be influenced by the sample shape. Corrections should therefore also be performed on the noise measurements to obtain the ``intrinsic" noise response. We are not aware of the correct procedure for this, although we stress that, while corrections change the scale of the response, they do not remove information and the important thing is the measured magnetic noise should scale with the measured susceptibility as first pointed out in Ref.~[\onlinecite{Reim86}].
To estimate the effects of demagnetization corrections on the noise, one can compare $D_i(f)$, obtained from the corrected $\chi_i''(f)$ susceptibility with the measured noise \cite{supmat}.
We note that at low temperature, typically below 500~mK, these corrections become almost negligible due to the very small value of $\chi'$ and are thus not relevant in the range where FDR violations are observed. These results also point to the difficulty of making quantitative comparisons between noise from experiments and from models in which the intrinsic response is simulated using periodic boundaries and the Ewald summation \cite{Samarakoon22, Hallen22}.

In summary, by measuring the FDR for \dto\ and \hto\ deep into the highly correlated spin ice regime,  we show that the FDR is satisfied in a large frequency range in both systems, including in the nonergodic region below 650 mK. We have confirmed the existence of two distinct relaxation times in \hto, which could be associated with the two tunnelling relaxation times proposed theoretically \cite{Tomasello19}. 
Below 300 mK and 400 mK in \dto\ and \hto, respectively, a clear violation to the FDR develops, with an excess noise appearing below 0.1 Hz in both compounds, leading to a fluctuation-dissipation ratio larger than 1 at 0.01 Hz. Such a deviation can only be explained by the presence of several timescales in the system. This violation of the FDR may be the signature of the sensitivity of the system to the Pauli states bandwidth. It is observed in a regime where the ac susceptibility possesses some low intensity residual relaxation processes and presents aging properties. 
To further characterize these out-of-equilibrium processes and tune the density of monopoles in the low temperature state, an important step forward would be to probe the FDR after controlled quench protocols, which is especially challenging for this kind of experiment. Our study also opens the way to new theoretical developments to understand the nature of the novel relaxation processes in this regime.


\section{Acknowledgments}
This work was supported by ANR, France, Grants No. ANR-15-CE30-0004 and No. ANR-19-CE30-0040. F. Morineau acknowledges financial support from the LANEF PhD Program. We are particularly grateful  to Gr\'egory Garde, Anne G\'erardin, Gilles Pont and Olivier Tissot for their technical support in the development of the noise magnetometer.  We acknowledge Ludovic Berthier, Steven Bramwell, Claudio Castelnovo, Bruno Tomasello and Mike Zhitomirsky for helpful discussion.  The crystal growth of \hto\ was performed by using facilities of the Materials Design and Characterization Laboratory in the Institute for Solid State Physics, the University of Tokyo under the Visiting Researcher's Program. The crystal growth of \dto\ was carried out at the University of Warwick, UK, and the work was funded by EPSRC, UK through Grant EP/T005963.

\bibliography{biblio}

\begin{thebibliography}{40}
\bibitem [{sup()}]{supmat}%
  \BibitemOpen
  \href@noop {} {}\bibinfo {howpublished} {See suplemental material, which
  includes (i) a description of the experimental set-up and samples, (ii)
  additional notes on the noise and ac susceptibility analysis}\BibitemShut
  {NoStop}%
\end{thebibliography}

\clearpage

\renewcommand{\thefigure}{S\arabic{figure}}
\renewcommand{\theequation}{S\arabic{equation}}
\renewcommand{\thetable}{S\arabic{table}}

\setcounter{figure}{0}
\setcounter{equation}{0}
\setcounter{table}{0}

\onecolumngrid
\begin{center} {\bf \large Supplemental Information : \\ Satisfaction and Violation of the Fluctuation-Dissipation Relation in spin ice materials} \end{center}
\vspace{1cm}
\twocolumngrid

\section{Methods}

\subsection{Noise and AC-susceptibility SQUID magnetometer}
\begin{figure}
\includegraphics[width=9cm]{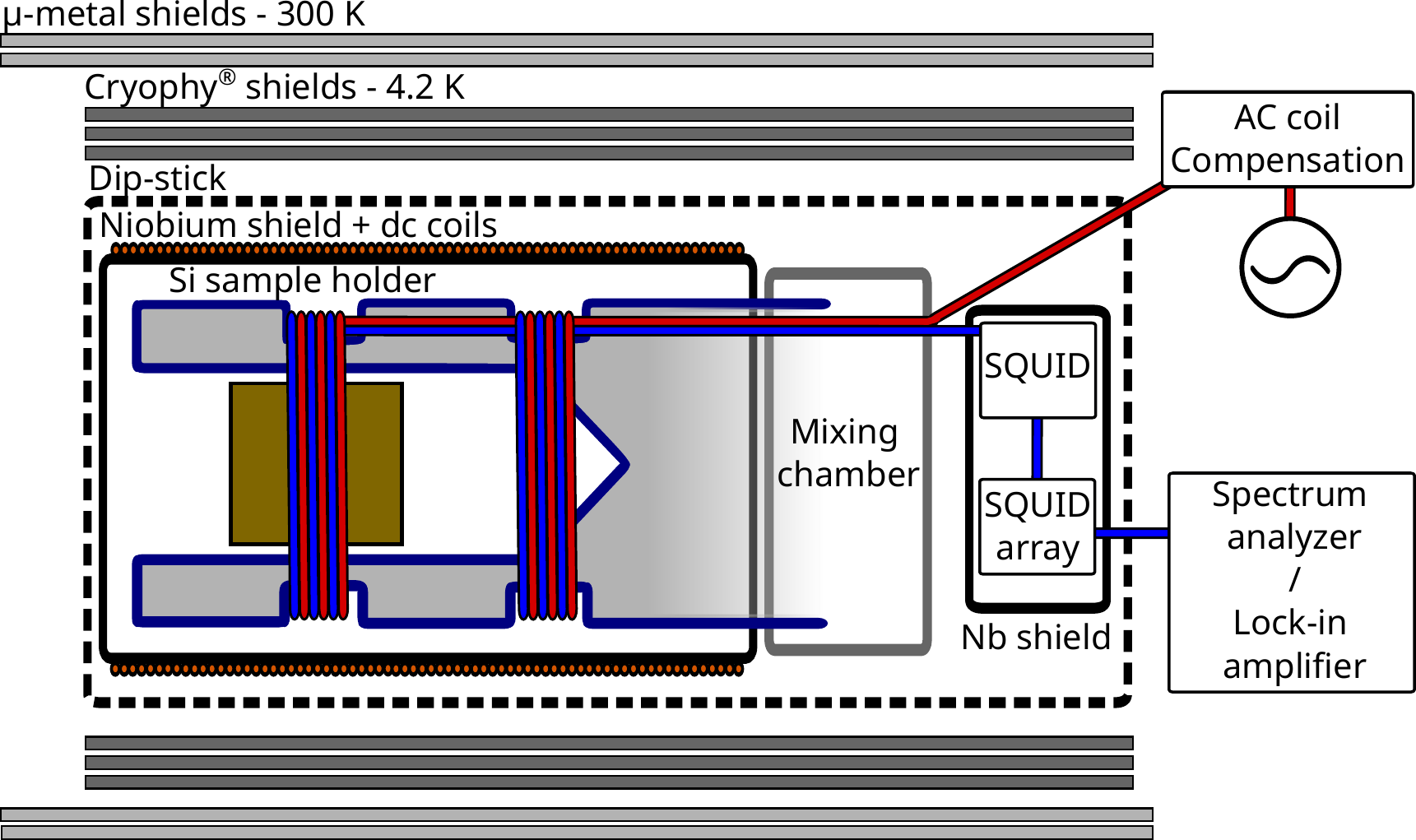}
\caption{\label{schema}Schematic of the experimental setup used to measure both magnetic noise fluctuations and the ac susceptibility. The gradiometer is represented in blue coils, and the excitation coils are represented in red. The sample is represented by a brown square inserted in the silicium sample holder.}
\end{figure}

The SQUID magnetometer setup (Figure \ref{schema})  was designed to measure both weak spontaneous noise signals and low-field ac-susceptibility in the same experimental environment. It ensures a minimal external residual magnetic flux in the sample area through the use of several layers of magnetic shielding. Outside the cryostat at room temperature two $\mu$-metal shields are used, and inside the helium bath, three Cryophy\textsuperscript{\textregistered} (a Ni-Fe-Mo alloy treated for low temperature performance) shields are inserted leading to a residual magnetic field of the order 50 nT in the bath. 

A miniature dilution refrigerator of the dip-stick variety carrying the probe is inserted into this environment. The probe itself is shielded by a Niobium superconducting shield, and consists of a superconducting pick-up coil wrapped around a cylindrical single crystal silicon sample holder attached to the mixing chamber, which ensures a good thermal contact and a low magnetic noise contribution.
If an external field is present in the sample environment, a dc Magnetization vs Temperature (MvsT) response will be measurable. The greater the external field the greater the MvsT response, which will be a source of spurious low frequency noise if the temperature is not perfectly stable. Thus in order to further reduce the residual magnetic field a dc coil was wrapped around the Niobium shield for additional field cancellation. Before the start of an experimental run, measurements of M vs T in the residual field were made. Then by heating the shield above its superconducting critical point with a small field in the dc coil, and then cooling the shield, the residual field could be reduced.  By repeating the procedure the field could be adjusted so that the MvsT response of the sample was nearly zero. The final fields were typically less than 5 nT in the sample area. 

The pick-up coil is a first order superconducting gradiometer composed of two coils wrapped in opposite directions, with the sample positioned in the center of one of the coils. This gradiometer allows us to measure the magnetic flux contribution from the sample via magnetic induction while canceling out first-order residual magnetic flux. The gradiometer is connected via mutual induction to the input coil of a Magnicon SQUID with a SQUID array preamplifier, which acts as a very low noise amplifier, measuring spontaneous magnetic fluctuations down to 1.6 $\mu \Phi_{0}/\sqrt{{\rm Hz}}$. The SQUID is thermalized at liquid helium temperature, ensuring that its setup point remains stable throughout the entire measurement. 

Inside the sample environment, two excitation coils in a series are wrapped to apply small ac magnetic fields (typically 50 $\mu$Oe peak to peak) to the sample. The sample response is then detected by the pick-up coil and processed with an SR830 Lock-in Amplifier, giving the ac susceptibility $\chi=\chi '-i\chi ''$, where $\chi'$ is the in-phase response and $\chi''$ the out-of-phase response of the sample. When the sample is cooled down to low temperature (typically under 500 mK), the ac susceptibility signal becomes really weak. It is thus necessary to use an adjustable current divider put between the two excitation coils to minimize any offset in $\chi'$ in order to reach the required sensitivity. Ac susceptibility is measured for each temperature from 0.001 Hz to 10$^{5}$ Hz. 

The noise power spectral density (PSD) is acquired by performing a Fast Fourier Transform (FFT) on the temporal spontaneous noise signal using an HP-35655A spectrum analyzer. Several frequency spans are measured sequentially to cover the entire measurable frequency range (0.002 Hz to 10$^{5}$ Hz). For each temperature, at least three measurements are averaged to improve the statistical accuracy and reduce the standard deviation of the measured noise.

\subsection{Calibration}
A calibration of the magnetometer was made by measuring the ac susceptibility of Pb spheres resulting in a calibration factor $k_{\rm Pb}$, expressed in emu$/\Phi_{0}$. However the calibration factor is sensitive to the sample position when it is not perfectly centered in the lower detection coil. In addition, when the sample size is comparable to or greater than the geometry of the pick-up coil, the effective measured volume of the sample is smaller than the actual sample volume which in turn affects the calibration.

In order to alleviate these problems, a scaling of the ac susceptibility (for each sample) measured in the set-up described above (in $\Phi_{0}$) was made to that measured (for each sample) in a homemade SQUID magnetometer. This magnetometer can operate down to 70 mK and makes measurements by the extraction method giving absolute values of the magnetization or susceptibility  in emu. The scaling was typically made at 4.2 K and 800~mK in order to have a robust real and imaginary susceptibility. A scaling factor $k$, expressed in emu$/\Phi_{0}$, is obtained and is typically close to $k_{Pb}$. 

The noise (in $\Phi^{2}_{0}/$Hz) is then scaled using the same $k$ factor with the following expression $S(({\rm emu}/{\rm cm}^{3})^{2}/{\rm Hz})=k^2/V^2S(\Phi^{2}_{0}/{\rm Hz})$, with $V$ the sample volume. The calibration is checked using the fluctuation-dissipation relation in the thermodynamic equilibrium regime from 4.2 K to 700 mK. For small samples we found that the relation is obeyed with no free parameters over the entire span.  However for large samples a mismatch between $S(f)$ and $D(f)$ was observed, and was corrected with an additional $k'$ factor, which accounts for the effective volume of sample measured. For the \hto\ sample, this factor is 2 (the sample is large  compared to the coils size) and for the \dto\ sample, this factor is around 2.5 (the sample is larger than the \hto\ one). The two factors, $k$ and $k'$, are fixed for a given measurement series but need slight adjustments for each new measurement campaign since they depend on the position of the sample within the pick-up coil system. 

\begin{figure}[h!]
\includegraphics[width=4cm]{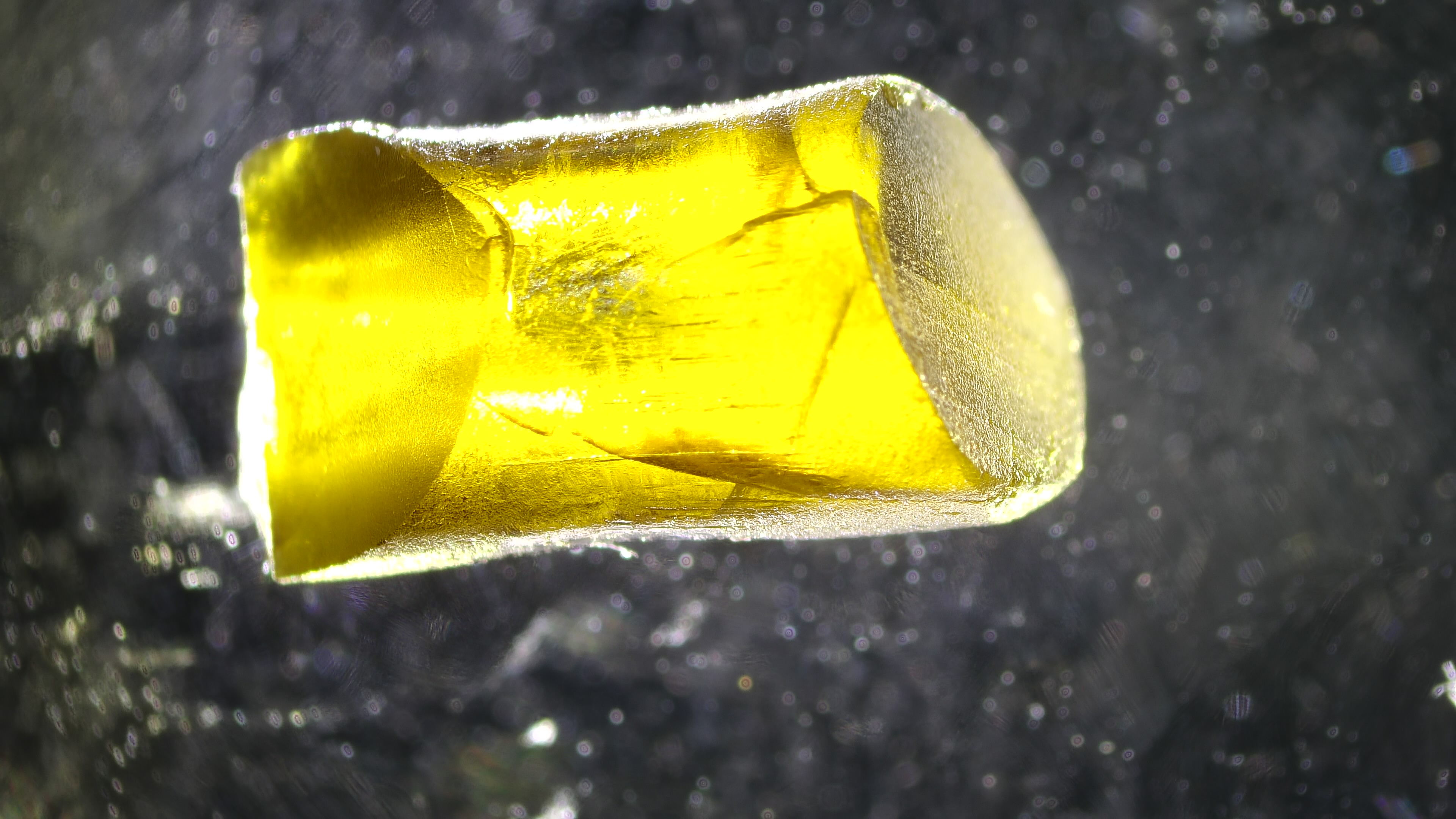}
\includegraphics[width=4cm]{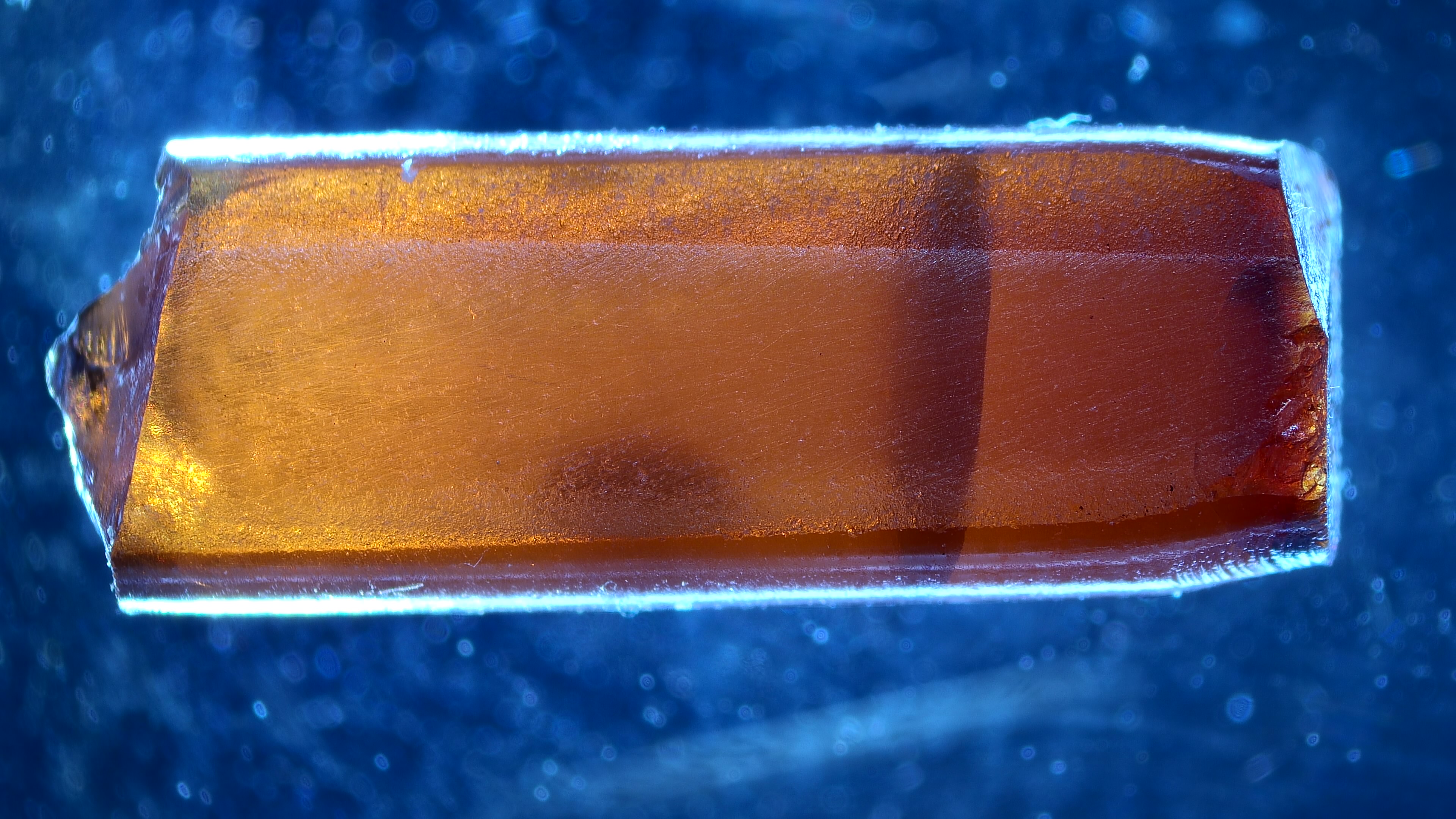}
\caption{\label{samples} Left photo : \hto\ sample. Right photo : \dto\ sample. }
\end{figure}

The two samples are shown on Figure \ref{samples}. The \hto\ sample (left) was prepared at the ISSP Tokyo by the floating-zone method using an infrared furnace equipped with four halogen lamps and elliptical mirrors. The crystals were grown under O$_2$ gas flow to avoid oxygen deficiency. The typical growth rate was 4 mm.h$^{-1}$. The \dto\ sample (right) was grown in Warwick, and was already measured in Refs. \onlinecite{Fennell04,Giblin18}.

\section{Noise and ac susceptibility analysis}
\subsection{Noise}

Before analyzing the noise curves, the background noise PSD and harmonics peaks resulting from external perturbations (such as turbopump rotation at 3 kHz) are removed from the measured data.

\begin{figure}[h!]
\includegraphics[width=8cm]{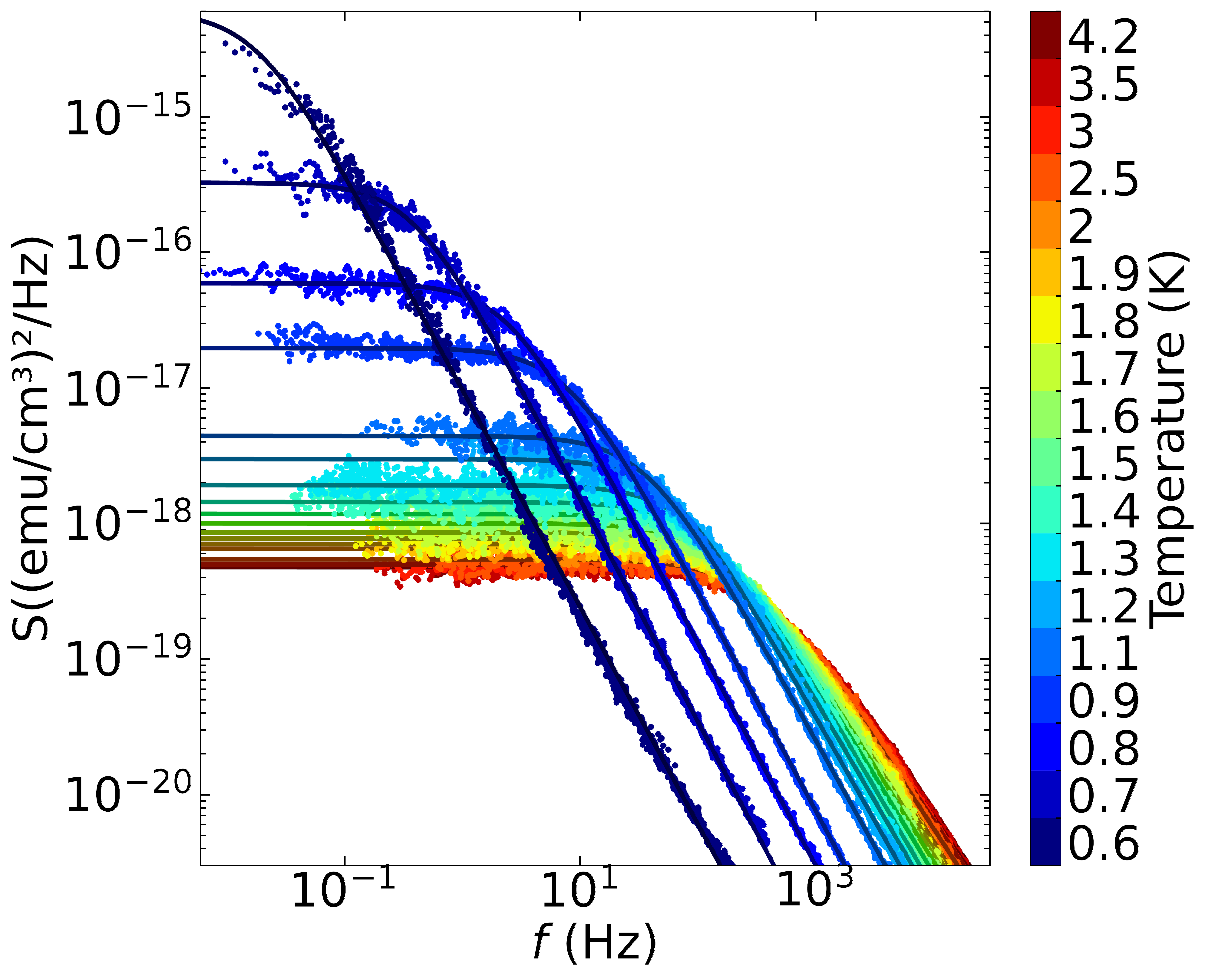}
\caption{\label{Dynonuk_noise_fit} Noise spectral density for \dto\ $S(f)$ on a logarithmic scale. Small dots represent measured data and solid lines represent the fits according to Equation (\ref{fit_S}).} 
\end{figure}

The \dto\ magnetization noise spectra are fitted using a least square cost function to the following empirical equation (Figure \ref{Dynonuk_noise_fit}) : 
\begin{equation}
S(f)=\frac{S_0}{1+(2\pi f \tau)^{\alpha}}
\label{fit_S}
\end{equation}

\begin{figure}[h!]
\includegraphics[width=5cm]{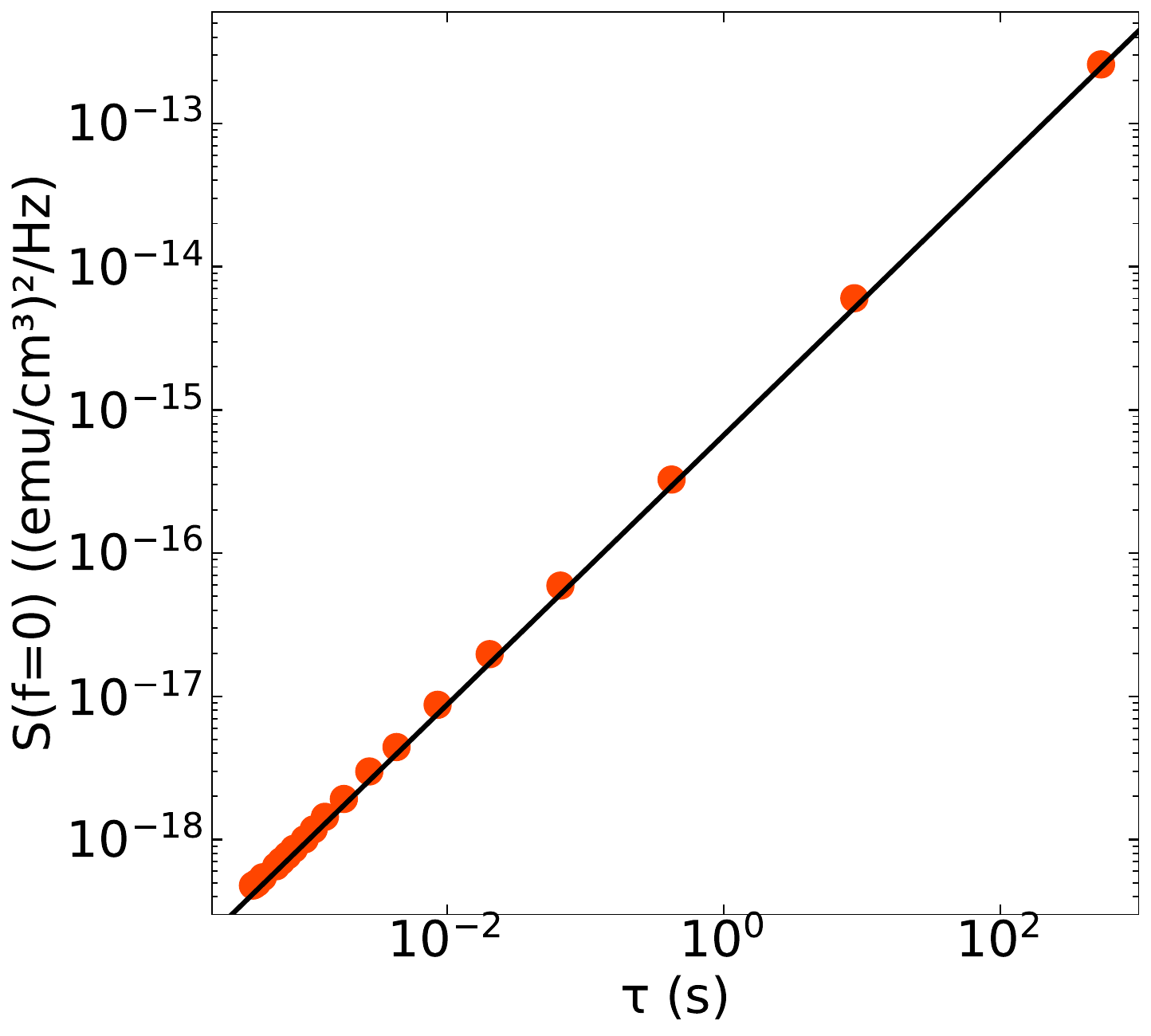}
\caption{\label{S0_nonuk}Noise at zero frequency $S_0$ as a function of $\tau$ for \dto. The solid black line is a fit to a power law with an exponent equal to 0.94.}
\end{figure}

Three free parameters are extracted from the fit: $S_0$ the noise at zero frequency (corresponding to the noise value on the plateau), $\tau$ the characteristic relaxation time of the system, and the power law exponent $\alpha$.

\begin{figure}[h!]
\includegraphics[width=8cm]{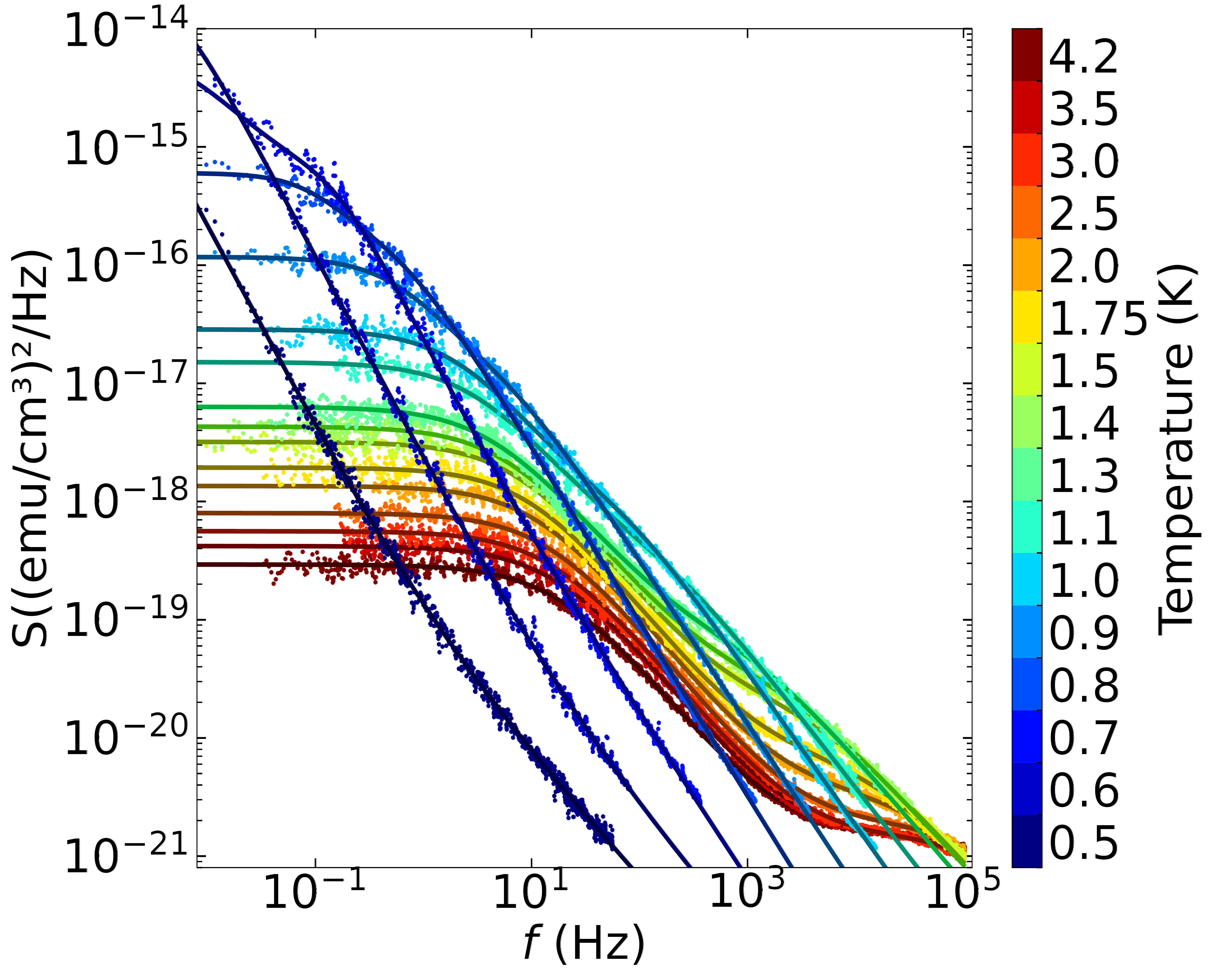}
\caption{\label{Ho_noise_fit} Noise spectral density for \hto\ $S(f)$ on a logarithmic scale. Smalls dots represent measured data and solid lines represent the fits according to Equation (\ref{fit_S_double}).}
\end{figure}

The dependence of $S_0$ on $\tau$ (see Figure \ref{S0_nonuk}) is almost linear, but is much better fitted with a power law with a 0.94 exponent, indicating deviations to the linear behavior predicted for a simple generation-recombination phenomenon.
\medskip

For the \hto\ noise spectra, two distinct relaxation times are noticeable. They are thus fitted with two sets of three free parameters, leading to the sum of two functions $S_1(f)$ and $S_2(f)$ as follows (Figure \ref{Ho_noise_fit}) : 
\begin{equation}
S(f)=\frac{S_{01}}{1+(2\pi f \tau_1)^{\alpha_1}}+\frac{S_{02}}{1+(2\pi f \tau_2)^{\alpha_2}}
\label{fit_S_double}
\end{equation}

\begin{figure}[h!]
\includegraphics[width=5cm]{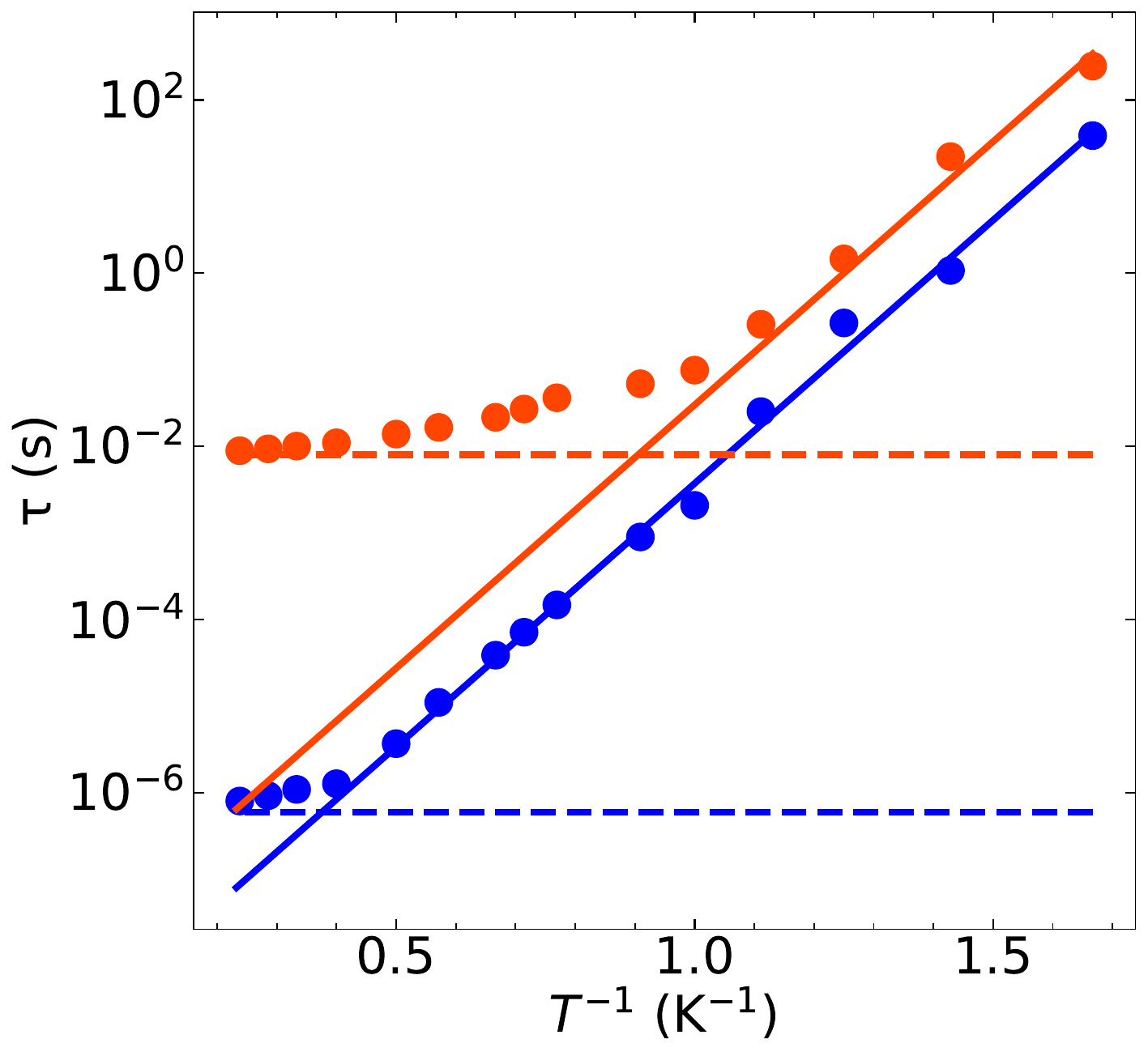}
\caption{\label{HTO_arrhenius} $\tau_{\rm slow}$ and $\tau_{\rm fast}$ vs $1/T$ in \hto. Dashed lines represent the tunneling times $\tau_{\rm t}$ obtained in the plateau region above 2.5 K, $\tau_{\rm t slow} = 8 \times 10^{-3}$ s and $\tau_{\rm t fast} = 6 \times10^{-7}$ s. Solid lines are Arrhenius laws $\tau_0 \exp(E/T)$ with $\tau_{\rm 0 slow} = 2.5 \times 10^{-8}$ s, $\tau_{\rm 0 fast} = 3.1 \times 10^{-9}$ s and the same energy barrier $E = 14$~K for both.}
\end{figure}

The two relaxation times in \hto\ have a similar temperature dependence on the whole temperature range (see Figure \ref{HTO_arrhenius}). They seem to correspond to two well separated tunneling processes which have a different single spin flip tunneling time depending on the local spin configurations with a similar activation energy.


\subsection{Ac susceptibility}

\begin{figure}[h!]
\includegraphics[width=8cm]{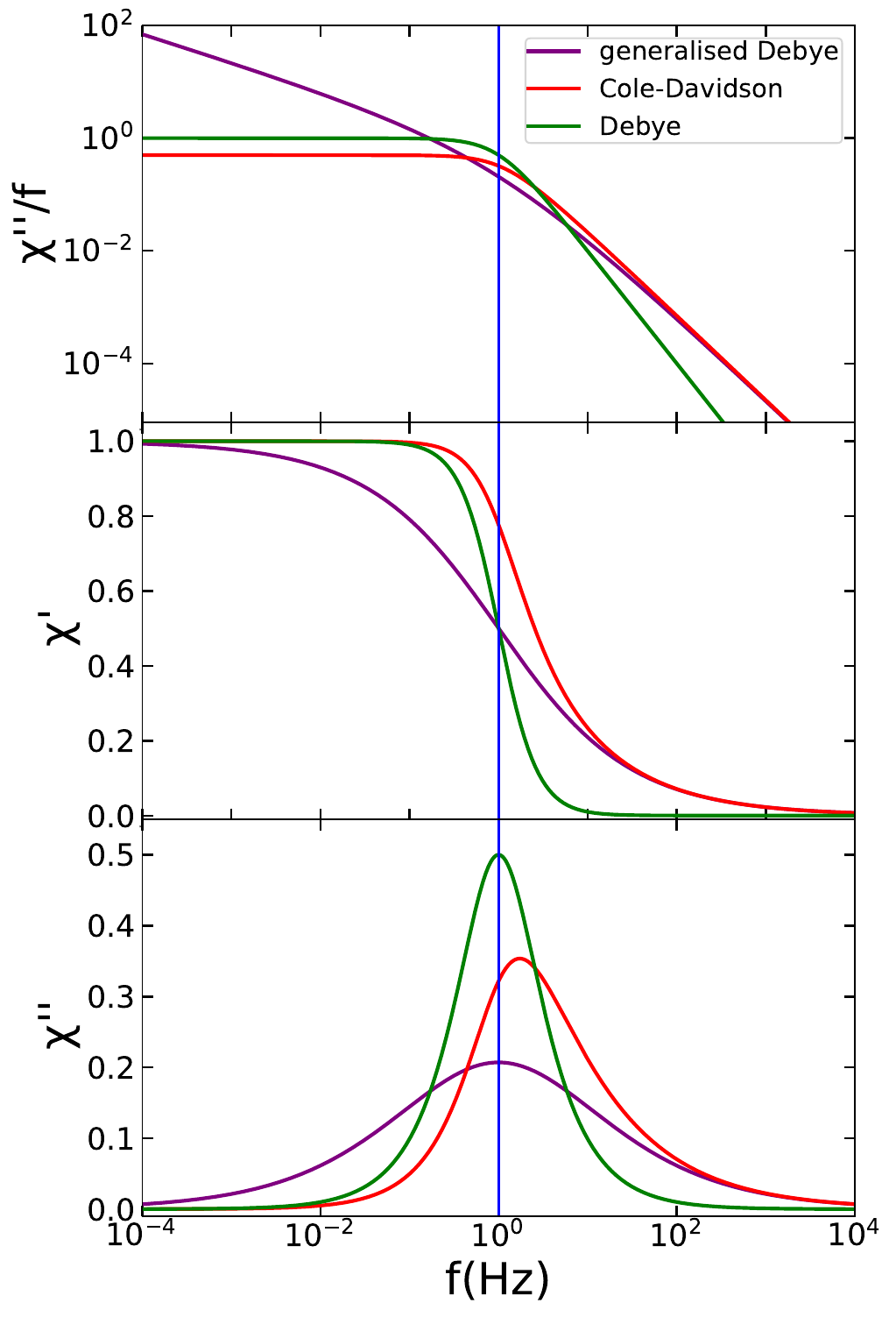}
\caption{\label{fit_model} Response functions as a function of frequency for the generalized Debye, the Cole-Davidson and the Debye model in purple, red and green respectively: $\chi"/f$, proportional to the noise PSD $S$ (top panel), real part $\chi'$ (middle panel) and imaginary part $\chi''$ (bottom panel) of the ac susceptibility. Parameters are: $\chi_{\rm S}$=0, $\chi_{\rm T}$=1, $\tau$=$\frac{1}{2\pi}$ (represented by the blue line) and $\beta$=$\gamma$=0.5.}
\end{figure}
Different empirical response functions are used to fit ac susceptibility measurements. The most straightforward response function one can use is introduced in the Debye model (Figure \ref{fit_model}, green curves):
\begin{equation}
\chi(f)=\chi_{\rm S}+\dfrac{\chi_{\rm T}-\chi_{\rm S}}{1+i2\pi f \tau}
\label{Debye}
\end{equation}

This expression describes phenomena that relax with a single time scale. However spin ice is a strongly correlated state in which this picture of a single characteristic time scale is insufficient. In this case, it is necessary to introduce a distribution of relaxation times with a characteristic time scale $\tau$. 

A model with a symmetric $\tau$ distribution can be obtained by introducing a new exponent parameter $\gamma$, giving the generalized Debye model (Figure \ref{fit_model}, purple curves): 
\begin{equation}
\chi(f)=\chi_{\rm S}+\dfrac{\chi_{\rm T}-\chi_{\rm S}}{1+(i2\pi f \tau)^\gamma}
\label{generalised Debye}
\end{equation}

Another empirical model, the Cole-Davidson model (Figure \ref{fit_model}, red curves) possesses an asymmetric $\tau$ distribution and is expressed as:
\begin{equation}
\chi(f)=\chi_{\rm S}+\dfrac{\chi_{\rm T}-\chi_{\rm S}}{(1+i2\pi f \tau)^\beta}
\label{Cole-Davidson}
\end{equation}

For the generalized Debye model, varying the exponent $\gamma$ from 1 to 0 broadens the distribution symmetrically around the $\tau$ value. It introduces a factor $(\omega\tau)^{\gamma}$ in the numerator of the $\chi''$ expression, resulting in a finite slope behavior at low frequency when considering the noise analog (see top panel, Figure \ref{fit_model}). This is incompatible with our noise measurements, which show a flat ``white noise" behavior in the low frequency limit. For the Cole-Davidson model, varying $\beta$ from 1 to 0 results in a strong asymmetric broadening for time scales shorter than $\tau$ due to a cutoff in the distribution of relaxation times. This cutoff results in a flat shape at low frequencies of the associated noise, which accords with the noise measurements.

\begin{figure}
\includegraphics[width=8cm]{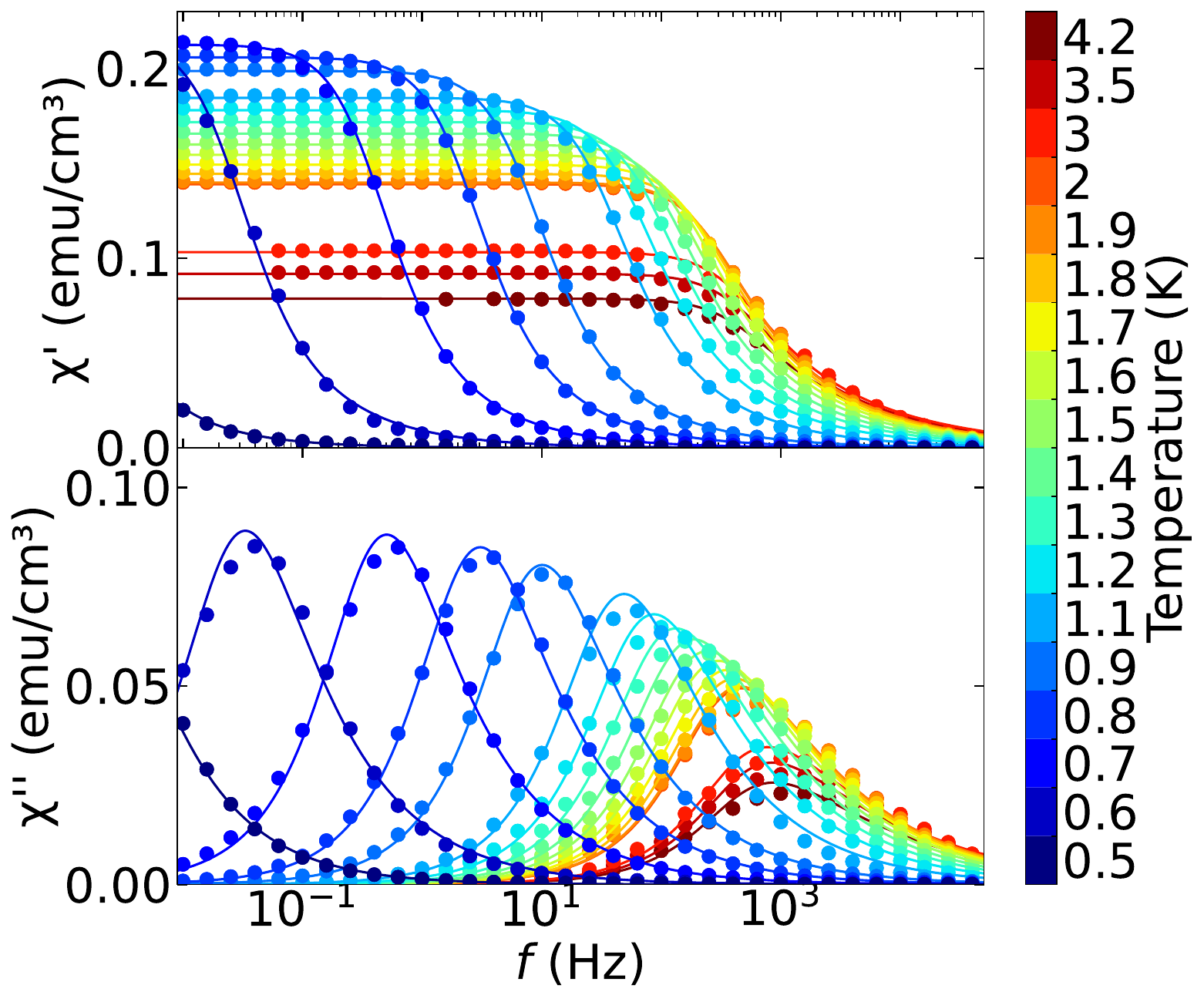}
\caption{\label{dto_acfit} Ac susceptibility $\chi'$ (top panel) and $\chi''$ (bottom panel) vs $f$ for different temperatures in \dto. Dots represent measured data and lines represent the Cole-Davidson fits. }
\end{figure}

Due to this constraint from the noise behavior, the Cole-Davidson model is chosen for the fitting procedure of the ac susceptibility curves. It is interesting to notice that the Cole-Davidson equation for $\chi''$ is not strictly identical to the equation used for the noise measurements, but both share the same overall features. The fit to the Cole-Davidson model is applied to both \dto\ and \hto. Similarly to noise fitting, the \hto\ fit function consists in the sum of two Cole-Davidson functions. $\chi'$ and $\chi''$ are fitted jointly to ensure that one set of parameters reproduces both quantities and obeys Kramers-Kronig relations. This analysis accurately describes the \dto\ data  (Figure \ref{dto_acfit}) across the entire temperature range. 

\begin{figure}[b]
\includegraphics[width=7.5cm]{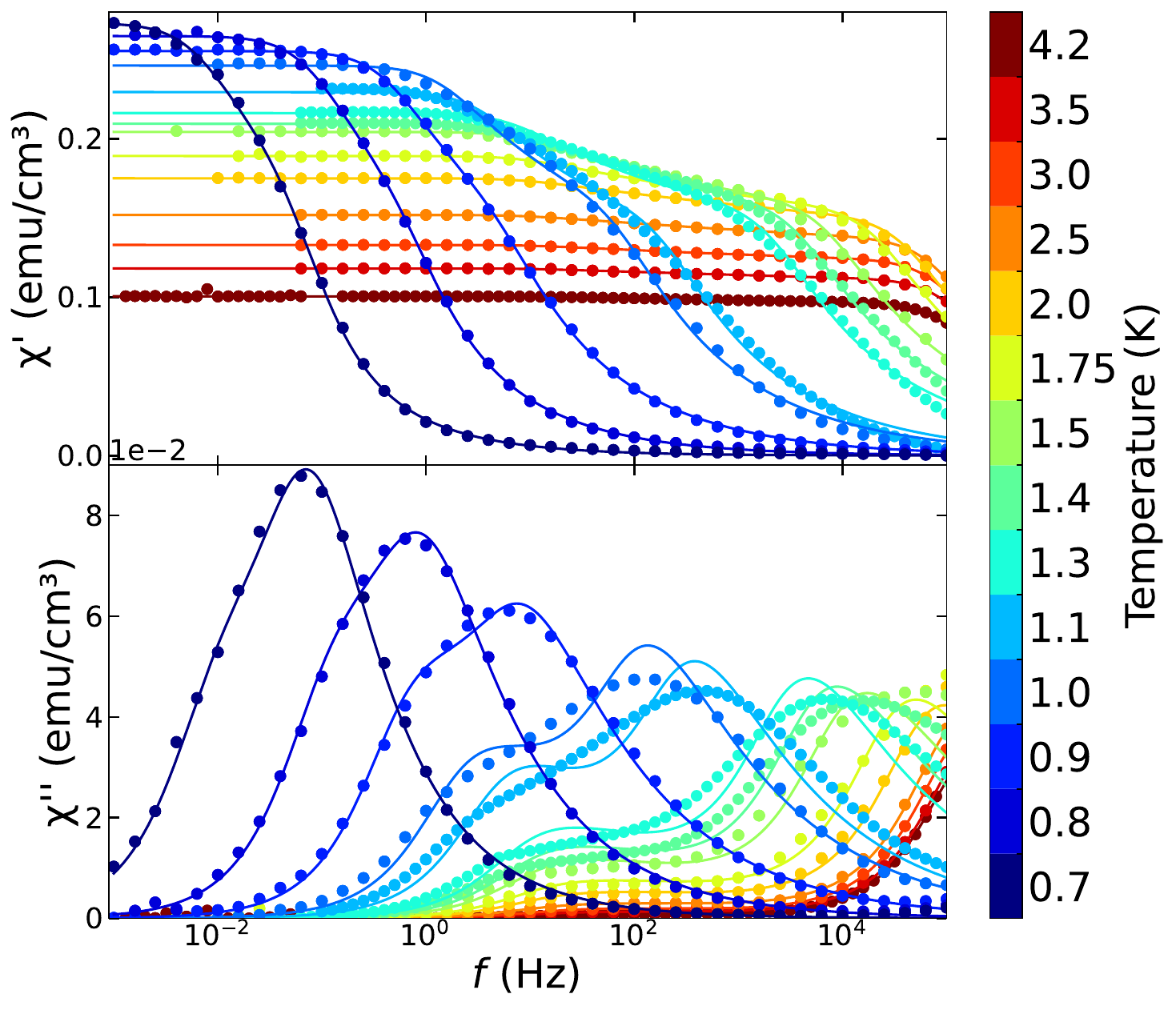}
\caption{\label{hto_acfit} Ac susceptibility $\chi'$ (top panel) and $\chi''$ (bottom panel) vs $f$ for different temperatures in \hto. Dots represent measured data and lines represent the Cole-Davidson fits.} 
\end{figure}

\begin{figure}[!]
\includegraphics[width=7.5cm]{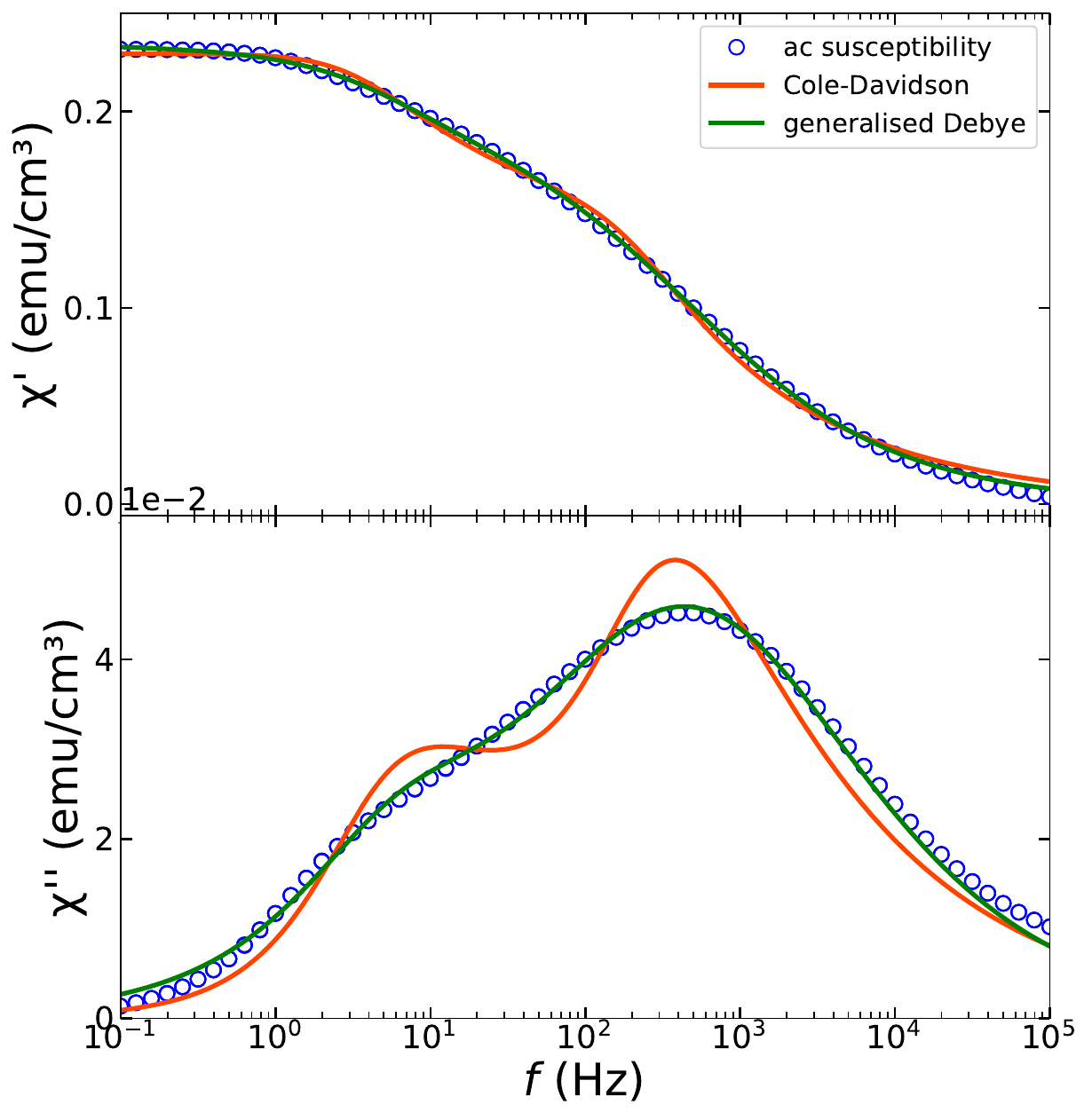}
\caption{\label{fit_1.1K_diff} Real part $\chi'$ (top panel) and imaginary part $\chi''$ (bottom panel) of the measured ac susceptibility in \hto\ at 1.1 K (blue dots) compared to Cole-Davidson (in orange) and generalized Debye (in green) fits.}
\end{figure}

In the case of \hto\, the fits are of a poorer quality in the intermediate temperature range (between 1 and 1.5~K) when the two time scales are well separated (Figure \ref{hto_acfit}). This difficulty is explained by the asymmetric behavior of the model, which induces a non vanishing high frequency contribution for a given broadness. Thus, the low frequency peak in the fit possesses a significant intensity at higher frequencies, which affects the fitting of the higher frequency peak. The fits then need to find an optimum between matching the broadness and the intensities. For the specific case of $T=1.1$ K, Figure~\ref{fit_1.1K_diff} shows that the generalized Debye model does not have this issue. However, for the reasons explained above, this generalized Debye model is incompatible with the noise measurements. These results nevertheless show that the ``simple" above equations cannot capture the dynamics of spin ice in the whole temperature and frequency range, and are just approximations to apprehend the dependence of the relaxation times in the system.


\begin{figure}
\includegraphics[width=8cm]{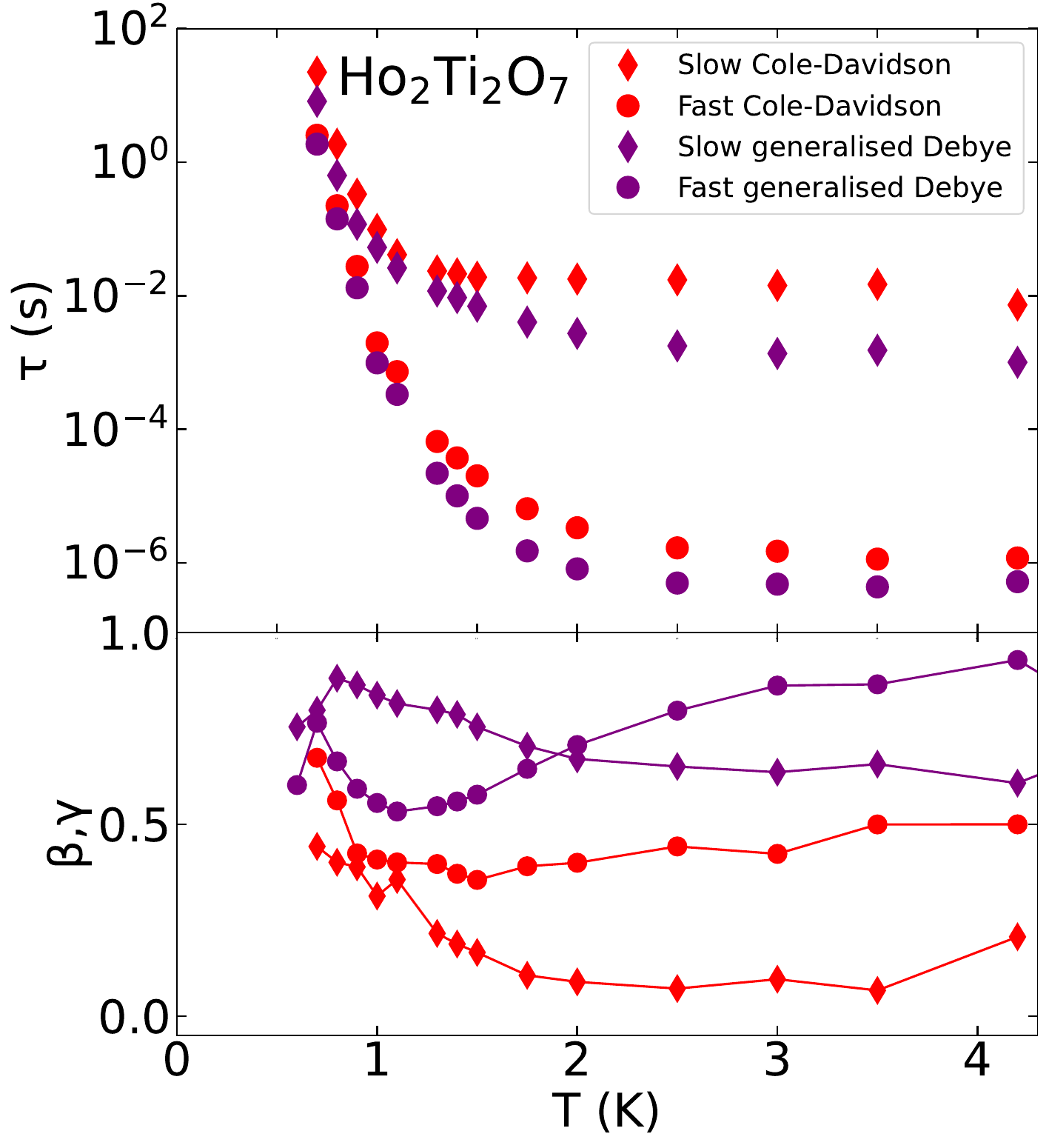}
\caption{\label{tau_alpha_CD_debye_diff} Relaxation times $\tau_{\rm slow}$ and $\tau_{\rm fast}$ (top panel) and exponents $\beta$ and $\gamma$ (bottom panel) in \hto\ obtained from ac susceptibility measurements. Blue symbols correspond to a Cole-Davidson model analysis and purple symbols correspond to a generalized Debye analysis.}
\end{figure}

In order to compare the influence of the fit model on the obtained parameters, we have performed the fitting procedure for the ac susceptibility for both generalized Debye and Cole-Davidson model. Figure \ref{tau_alpha_CD_debye_diff} shows that the overall behavior of the exponent parameters and the relaxation times is similar with both models. It is interesting to stress out that the Cole-Davidson model shifts the $\tau$ values to longer times compared to the generalized Debye model. This is due to the asymmetry of the distribution function of the Cole-Davidson model which results in a maximum in $\chi$'' located at a higher frequency for a same $\tau$ parameter (Figure \ref{fit_model}, bottom panel).

\subsection{AC susceptibility aging effect}
\begin{figure}[h!]
\includegraphics[width=8cm]{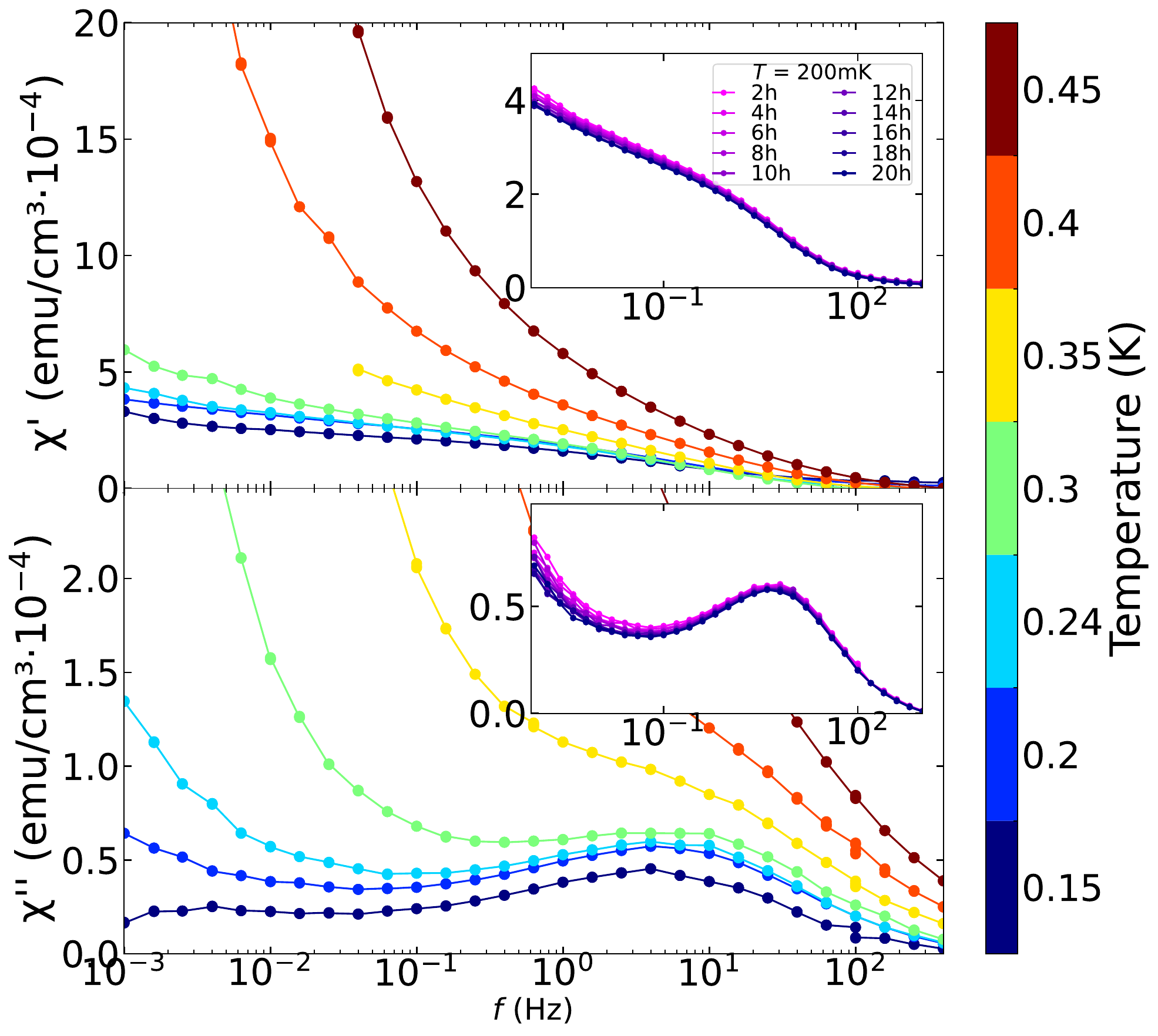}
\caption{\label{Dynonuk_lowT} ac susceptibility $\chi'$ (top panel) and $\chi''$ (bottom panel) vs $f$ for different temperatures below 300 mK in \dto. Insets : Susceptibility measured at 200 mK after different waiting times. }
\end{figure}

\begin{figure}[h!]
\includegraphics[width=6.5cm]{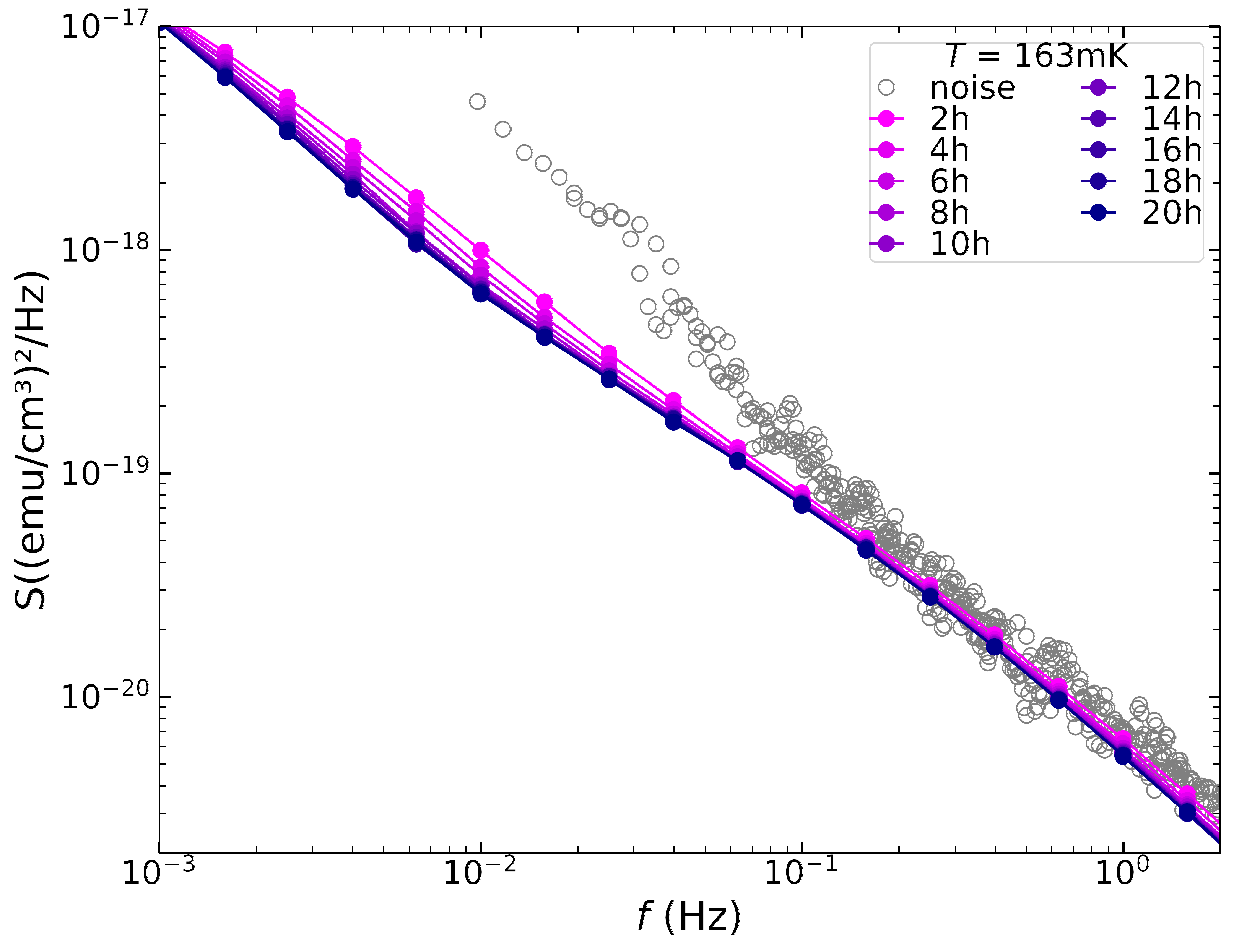}
\caption{\label{noise_aging_effect}FDR plot in \hto: $S(f)$ (grey empty dots) and $D(f)$ (dots) after different waiting times at 163 mK. }
\end{figure}

When cooling down to temperatures below 400 mK, we observe a new dissipative process in the ac susceptibility in both \dto\ and \hto. It is shown here in Figure \ref{Dynonuk_lowT} for \dto, and in Figure 5 of the main text for \hto. This dissipative process is observed below 10 Hz in \dto\ and 1 Hz in \hto. This dissipative process is associated to an aging effect, which appears to be weaker in \dto\ than in \hto\ (see insets of Figure \ref{Dynonuk_lowT}).

Figure \ref{noise_aging_effect} shows how this aging effect relates to the noise fluctuation spectrum at 163 mK in \hto. Above 0.1 Hz (which corresponds to the frequency below which the FDR is not obeyed), the relaxation of the ac susceptibility remains within the range of the scattering of the noise data. The aging effect is strongest below this frequency, with an equilibration time of 14 hours in \hto\ at 163 mK and of 10 hours in \dto\ at 200 mK.

\subsection{Demagnetization correction effects}

\begin{figure}[b]
\includegraphics[width=4.2cm]{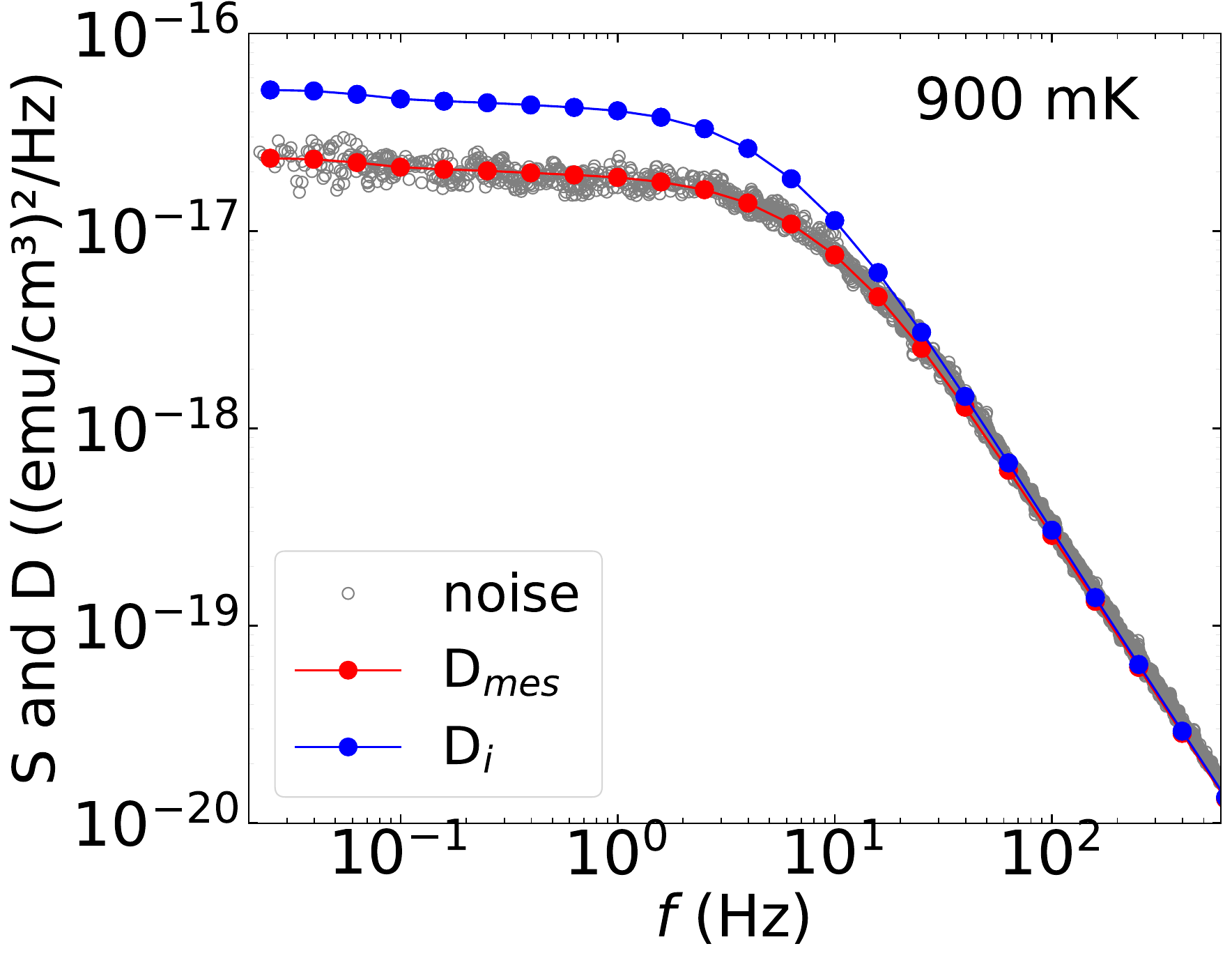}
\includegraphics[width=4.2cm]{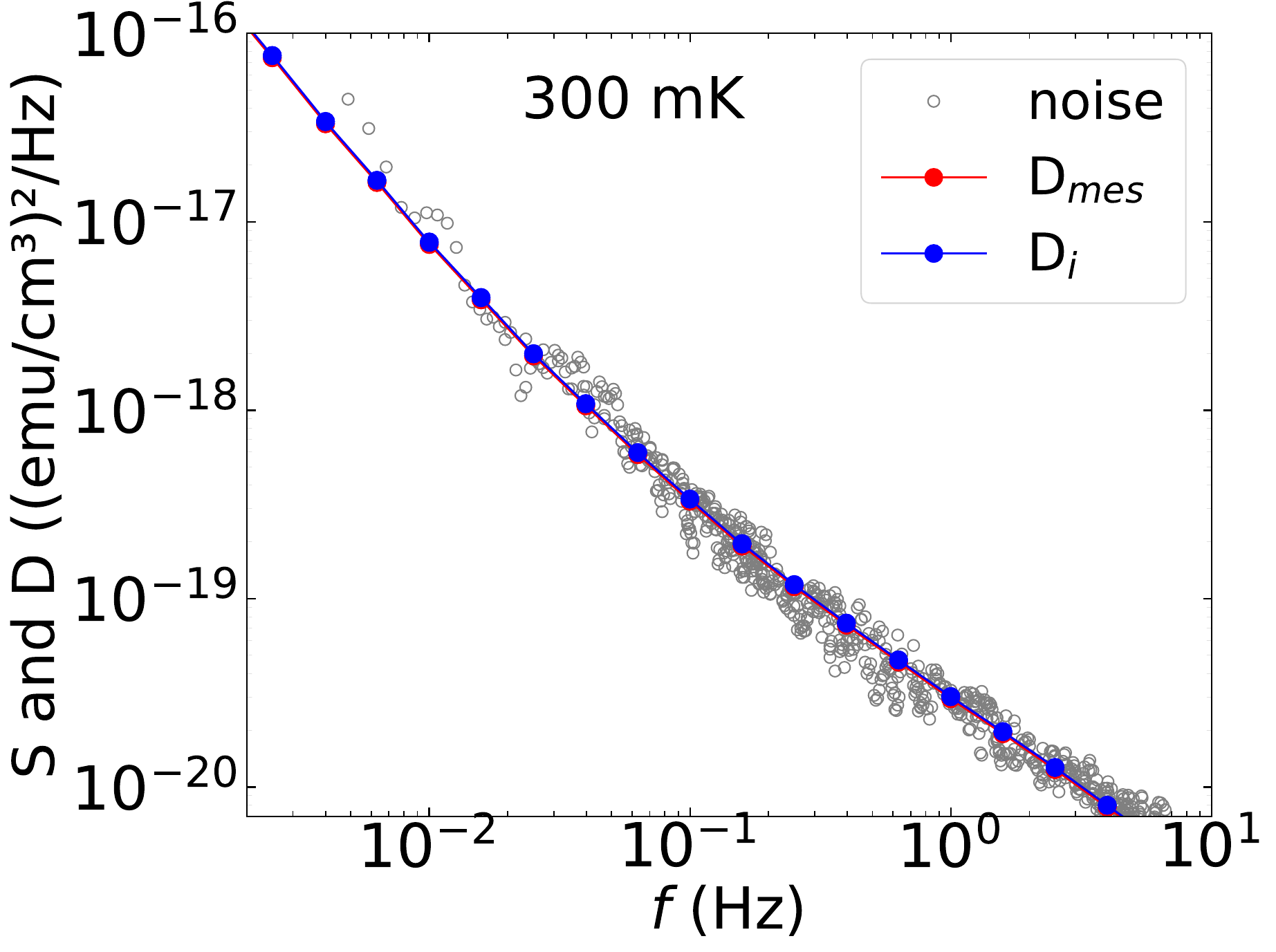}
\caption{\label{demag_effect}FDR plot at 900 mK (left panel) and at 300 mK (right panel) in \dto: $S(f)$ (grey empty dots) and $D(f)$ (red dots for no applied corrections and blue dots when correction is applied). }
\end{figure}

 {\color{black} Figure \ref{demag_effect} shows an example on \dto\ of the effect of the N demagnetizating factor on the dissipation part $D(f)$ of the FDR at two temperatures. The corrections on $\chi$ are performed using Equation \ref{demag} with an estimated N factor of 1.65 in cgs unit. $D_i(f)$ is then deduced using $\chi''_i$ instead of $\chi''_{mes}$ in the FDR. 

\begin{equation}
\label{demag}
\begin{split}
\chi'_{i}=\frac{\chi'_{mes}-N(\chi'^{2}_{mes}+\chi''^{2}_{mes})}{(1-N\chi'_{mes})^{2}+(N\chi''_{mes})^{2}} \\
\chi''_{i}=\frac{\chi''_{mes}}{(1-N\chi'_{mes})^{2}+(N\chi''_{mes})^{2}}
\end{split}
\end{equation}

The measured $D(f)_{mes}$ and $S(f)_{mes}$ are expected to be reduced in amplitude and compressed in time compared to the intrinsic ones. This effect depends strongly on the intensity of the ac susceptibility and becomes small under 600 mK. When the spin system is at equilibrium, the expected intrinsic magnetic noise should match through the FDR to the blue curve of $D_{i}(f)$ meaning corrections also needs to be applied to the measured noise $S_{mes}(f)$. However, no simple corrections can be applied to $S_{mes}(f)$ to recover the expected intrinsic noise.
}



\end{document}